\documentclass[10pt]{article}
\usepackage{flafter,amsmath,amssymb,latexsym,psfrag,graphicx,color,indentfirst}
\usepackage[margin=2.5cm]{geometry}
\usepackage[numbers,sort&compress]{natbib}

\usepackage[colorlinks,
linktocpage=true,
linkcolor=black,
citecolor=black
]{hyperref}

\newtheorem{theorem}{Theorem}[section]
\newtheorem{defn}[theorem]{Definition}

\newtheorem{prop}[theorem]{Proposition}
\newtheorem{example}[theorem]{Example}

\allowdisplaybreaks

\makeatletter

\@addtoreset{figure}{section}
\makeatother

\begin{document}
	\setlength\arraycolsep{2pt}
	\date{\today}
	
	\title{Wave Propagation in Pure-Time Modulated Step Media\\ With Applications to Temporal-Aiming}
	\author{Mourad Sini$^1$, Haibing Wang$^{2,3}$, Qingyun Yao$^4$
		\\$^1$RICAM, Austrian Academy of Sciences, A-4040, Linz, Austria \qquad\qquad\quad\quad
		\\E-mail: mourad.sini@oeaw.ac.at
		\\$^2$School of Mathematics, Southeast University, Nanjing 210096, P.R. China
		\\$^3$Nanjing Center for Applied Mathematics, Nanjing 211135, P.R. China\qquad
		\\$^4$School of Mathematics and Statistics, Nanjing University of Information\qquad
		\\ Science and Technology, Nanjing, 210044, P.R. China \qquad\qquad\qquad\qquad\qquad
		\\E-mail: hbwang@seu.edu.cn, qyyao@nuist.edu.cn
	}
	
	\maketitle
	\begin{abstract}
		
		We analyze the propagation of an incident electromagnetic wave in a purely-time modulated medium. 
		Precisely, we assume that the permeability is unchanged while the permittivity has a multiple-step profile in time and uniformly constant in space. For this, we use time-dispersive Lorenz's model with time-dependent plasma frequency with highly concentrated values near the centers of the steps' intervals.  Under certain regimes linking the number of steps to the contrasts of the permittivity,
		we can generate effective permittivity having positive and high values on a finite interval of time which behaves as a 'wall' that kills the forward waves and keeps only the backward waves (i.e. enabling full reflection). But this happens if these high values are away from a discrete set. If these high values are close to the elements of this discrete set, then the effective medium behaves as a 'well' that absorbs all the forward waves and hence there is no backward waves (i.e. enabling full transmission). {\it{Such results are reminiscent to the 'wall' and 'well' effects known in the quantum mechanics theory}}.

		\bigskip
		{\bf Keywords:} Time-Dependent Media, Electromagnetism, Temporal Aiming, Effective Media, Wall and Well Effects.\\
		
		{\bf MSC(2010): } 35L05, 35C20.
		
	\end{abstract}

	\section{Introduction and statement of the results}\label{intro}
	\setcounter{equation}{0}
	\subsection{General introduction and motivation}
	
	When an incident wave propagates through a changed material (a discontinuous one for instance), {\it{in space}}, then the frequency of the reflected wave is the same as the incident one while the wavenumber changes. This change of wavenumber across the two adjacent materials has tremendous implications. For instance, injecting, in a background media, a cluster of small particles having moderate (\cite{Bensoussan, Challa-1, Cioranescu-Murat, Mazya-1, Mazya-16, Nieves-17, AM, AM-1, Jikov, Ramm-7, Ramm-15, Sini-Wang-1}) or highly (\cite{ACP-2, AFLYH, AH, Bouchitte-Schweizer, BBF, Chen-Lipton-2013, Cher-Cooper, Cher-Ersh-Kise, L-S:2016, Schweizer:2017, Suslina-2019}) contrasted material properties, as compared to the ones of the background, modifies its behavior in reaction to incident waves. We can distribute the cluster in such a way that the resulting material will reflect or absorb more (or less) energy of the incident wave, at will. An extreme situation is when the new material changes the sign producing $\epsilon$-negative or $\mu$-negative materials and even double negative ones (where both $\epsilon$ and $\mu$ are negative valued). Such materials are very attractive as we can use them for many applications in material sciences and imaging, in broader sense, (as for enhancing the image resolution).
	\bigskip
	
	When an incident wave propagates through a changed material {\it{in time}}, however, the frequency of the reflected wave is different from the used one while the wave number is unchanged. Due to the symmetry roles of time and space in the wave propagation, similar, or related, results as in pure-space media based material generation are expected. In particular, we set the following two goals.
	
	\begin{enumerate}
		\item  We can generate effective permittivity having extremely high values on a finite interval of time which behaves as a 'wall' that kills the forward waves and keeps only the backward waves. However, if we use time-dispersive Lorentz's model with time-dependent plasma frequency, see \cite{Caloz-Leger-2020}, then the effective permittivity will have negative values on the finite interval of time which allows it to behave as a 'well' that absorbs all the forward waves and hence there is no backward waves. {\it{This is reminiscent to the 'wall' and 'well' effects known in the quantum mechanics theory}}.
		
		\item Another expected result is the possibility of 'redirecting' waves. Indeed, if the permittivity profile changes from isotropic to anisotropic, then the forward wave will follow the orientation of this anisotropy which allows redirecting wave packages at will, \cite{Pena}. 
	\end{enumerate}
	Such properties which are related to the so-called {\it{temporal-aiming}} have important implications, see \cite{Caloz-Leger-2020, Galiffi-al-2022, Huidobro-et-al, Huidobro--Pendry, Lannebere-Morgado-Silveirinha, Pena}. 
	In this work, we focus on the first goal. Electromagnetic models assuming time-modulated (or in general space-time modulated) materials can be found in classical books as \cite{Kalluri-book, Lurie} but also in more recent works with modern and uptodate motivations, see the review papers \cite{Caloz-Leger-2020, Galiffi-al-2022} for instance. Here, we consider the Maxwell-Lorentz model 
	
	\begin{equation}\label{Drude-Maxwell-Model}
		\left\{~~~
		\begin{array}{rll}
			\epsilon_0 \frac{\partial \boldsymbol{E}}{\partial t}+\boldsymbol{J}-\nabla \times \boldsymbol{H}=0,\\
			\\
			\mu_0 \frac{\partial \boldsymbol{H}}{\partial t}+\nabla \times \boldsymbol{E}=0,\\
			\\
			
			\frac{\partial\boldsymbol{J}}{\partial t}-\epsilon_0\omega^2_p\boldsymbol{E}=0,
		\end{array}
		\right.
	\end{equation}
where $\boldsymbol{E}$ and $\boldsymbol{H}$ are the electric and magnetic field respectively, $\boldsymbol{J}$ is the free electron density and $\omega_p:=\omega_p(x, t)$ the plasma frequency. In the sequel, we take the plasma frequency to be solely time dependent, i,e. $\omega_p:=\omega_p(t)$. A motivation of the model above can be found in \cite{Kalluri-book}, chapter 1, and the references therein. A more general form of the third relation in (\ref{Drude-Maxwell-Model}) is given in the presence of static magnetic field $B_0$ as 
	\begin{equation}\label{more-general-form-J-E}
	\frac{\partial\boldsymbol{J}}{\partial t}+\boldsymbol{J}\times {\bf{\omega_b}}-\epsilon_0\omega^2_p\boldsymbol{E}=0
	\end{equation}
	where ${\bf{\omega_b}}:= \omega_b B_0$ and $\omega_b$ is the absolute value of the electron gyrofrequency, see \cite{Kalluri-book} chapter 6.
	With such a constitutive relation, we model an anisotropic plasma medium. Let us mention here that, contrary to the reduced model, the general one provides with an integro-differential model that features anisotropy which might be useful for our purpose of time-aiming with redirecting waves, i.e. the goals (1)-(2) above.  The mathematical analysis of this model is more challenging and it will be the object of the future study. In addition, in this current work, we assumed that $\omega_p$ is solely time dependent. It can also be space dependent. The particular situation where it is spacetime dependent but with support in small scaled domain, in space variables, is of great importance in modeling spacetime modulated nanoparticles. Similarly, one can consider ${\bf{\omega_b}}$ to be spacetime dependent as well. These situations are parts of ongoing research program that we are establishing in the framework of electromagnetic wave propagation in the presence of spacetime modulated particles.     
	
	\subsection{Statement of the results}\label{Statement-results}
	
	In the present work, we neglect this static magnetic field $B_0$ and rather use the reduced constitutive equation in (\ref{Drude-Maxwell-Model}) which reduces to an isotropic model and discuss the goal (1) above. The third relation in (\ref{Drude-Maxwell-Model}) allows the model above to be written only in the electric field:
	
	\begin{equation}\label{reduced-model}
		\epsilon_0 \frac{\partial^2 \boldsymbol{E}}{\partial t^2}+\mu_0^{-1} \nabla \times \nabla \times \boldsymbol{E}+ \epsilon_0 \omega^2_p(t)\boldsymbol{E}=0.
	\end{equation}

	We look for guided waves of the form \begin{equation*}
		\boldsymbol{E}(\boldsymbol{x},\,t):=e^{i \boldsymbol{k}\cdot \boldsymbol{x}} \boldsymbol{\widetilde{E}}(t) =e^{i \boldsymbol{k}\cdot \boldsymbol{x}} (\widetilde{E}_1(t),\,\widetilde{E}_2(t),\,\widetilde{E}_3(t))^T,\;~ \boldsymbol{k}:=(k_1,\,k_2,\,k_3)^T.
	\end{equation*}  
	
	Then with
	\begin{equation*}
		\nabla\times\nabla\times\boldsymbol{E}=-e^{i\boldsymbol{k}\cdot\boldsymbol{x}}\,\boldsymbol{k}\times\left(\boldsymbol{k}\times\boldsymbol{\widetilde{E}}\right)=-e^{i\boldsymbol{k}\cdot\boldsymbol{x}}\,\left[\left(\boldsymbol{k}\boldsymbol{k}\cdot\boldsymbol{\widetilde{E}}\right)-|\boldsymbol{k}|^2\boldsymbol{\widetilde{E}}\right],
	\end{equation*}
	we have 
	\begin{equation}\label{maxwell-changing-equation}
		\epsilon_0\,\partial^2_t \boldsymbol{\widetilde{E}}-\mu_0^{-1}\boldsymbol{k}\left(\boldsymbol{k}\cdot\boldsymbol{\widetilde{E}}\right)+(\mu_0^{-1}|\boldsymbol{k}|^2+\epsilon_0 \omega^2_p(t))\boldsymbol{\widetilde{E}}=0, \quad t\in\mathbb{R}.
	\end{equation}
	Taking the inner product of \eqref{maxwell-changing-equation} and $\boldsymbol{k}^\perp$, where $\boldsymbol{k}^\perp\cdot\boldsymbol{k}=0$ and $|\boldsymbol{k}^\perp|=1$, then $\widetilde{E}:=\boldsymbol{\widetilde{E}}\cdot\boldsymbol{k}^\perp$ satisfies the scalar equation 
	\begin{equation}\label{The-scalar-model}
		\partial^2_t \widetilde{E}(t)+(|\boldsymbol{k}|^2\,(\epsilon_0\mu_0)^{-1}+\omega^2_p(t))\,\widetilde{E}(t)=0.
	\end{equation}
	As incident wave, we take $\boldsymbol{\widetilde{E}_{inc}}(t):= \boldsymbol{k}^\perp e^{i|\boldsymbol{k}|\sqrt{(\epsilon_0 \mu_0)^{-1}}t}$ which satisfies  $$\partial^2_t \boldsymbol{\widetilde{E}_{inc}}(t)-(\epsilon_0\mu_0)^{-1}\boldsymbol{k}\left(\boldsymbol{k}\cdot\boldsymbol{\widetilde{E}_{inc}}\right)+|\boldsymbol{k}|^2(\epsilon_0\mu_0)^{-1}\boldsymbol{\widetilde{E}_{inc}}(t)=0.$$
	\bigskip
	Denoting $\widetilde{E}_{inc}:=\boldsymbol{\widetilde{E}_{inc}}\cdot\boldsymbol{k}^\perp=e^{i|\boldsymbol{k}|\sqrt{(\epsilon_0 \mu_0)^{-1}}t}$, then we rewrite (\ref{The-scalar-model}) as 
	\begin{equation}\label{maxwell-D-Dinc-kperp}
		\partial_t^2(\widetilde{E}-\widetilde{E}_{inc})+|\boldsymbol{k}|^2(\epsilon_0\mu_0)^{-1}(\widetilde{E}-\widetilde{E}_{inc})=-\omega_p^2(t)\widetilde{E}.
	\end{equation}
	\noindent We set 
	\begin{equation*}
		\kappa:=|\boldsymbol{k}| \sqrt{(\epsilon_0 \mu_0)^{-1}},
	\end{equation*}
	then assuming that the scattered field $\widetilde{E}^s:=\widetilde{E}-\widetilde{E}_{inc}$ satisfies the long-time  condition\footnote{Actually, we only need to impose the condition for $t<0$ and $t>T$.} 
	\begin{equation}\label{SRC}
		(\partial_{r}-i\kappa)\widetilde{E}^s(t)=0, \quad r:=|t|, \mbox{ large enough,}
	\end{equation} we derive the integral equation
	\begin{equation}\label{integral-representation}
		\widetilde{E}(t)-\int^{+\infty}_{-\infty}\omega_p^2(t)\Phi(t,\,s)\widetilde{E}(s)ds=\widetilde{E}_{inc}(t),\quad t\in\mathbb{R},
	\end{equation}
	where $\Phi(\cdot,\,\cdot)$ is the fundamental solution of the Helmholtz equation in $\mathbb{R}$ and $\Phi(t,\,s) =\frac{i}{2\kappa}e^{i\kappa|t-s|}$. 
	\bigskip
	
\begin{figure}[htp]
\begin{center}
	\includegraphics[width=0.75\textwidth,height=6cm]{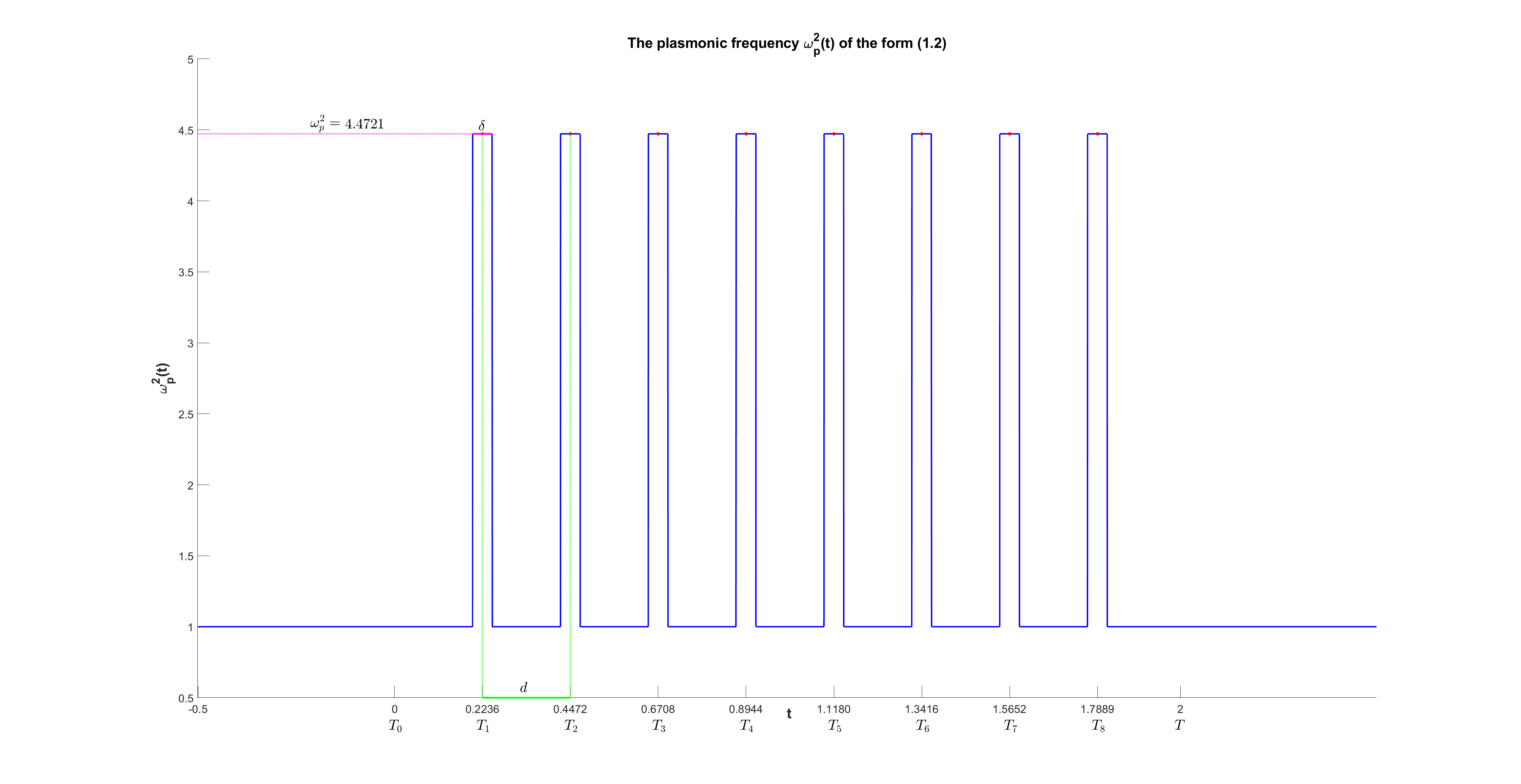}
	\caption{The model $\omega_p^2(t)$ with $C=1$, $h=0.5$  $\delta=0.05$, $l=0.5$ and $d=\delta^l$.}\label{epsilon}
\end{center}
\end{figure}

	In the following definition, we describe the parameters entering into \eqref{system-1} (or the equivalent model given by \eqref{The-scalar-model} and \eqref{SRC}).
	\begin{defn}\label{defn1} We assume the following properties.
		\begin{enumerate}
			\item We take the plasmonic frequency $\omega^2_p(t)$ of the following profile \footnote{We can take as the value $\omega_p^2(t)$ in $\mathbb{R}\setminus \cup^N_{j=1} I_j$ any constant $\widetilde{\omega}_p^2$, not necessary zero. In this case, we should replace $\kappa$ by $\sqrt{\kappa^2+\widetilde{\omega}_p^2}$.}
			\begin{equation}\label{material-time-variable}
				\omega^2_p(t)=0 \mbox{ in } \mathbb{R}\setminus \cup^N_{j=1} I_j
				\mbox{ and } \omega^2_p(t)=C\delta^{-h} \mbox{ in } I_j:=(T_j-\delta/2,\,T_j + \delta/2),\ j=1,\,2,\,\dots,\,N,
			\end{equation}
			where $C$ is a positive constant and $h\in(0,\,1]$. See Figure \ref{epsilon} for an illustration. 
			\noindent Let  $[T_1-\frac{\delta}{2},\,T_N+\frac{\delta}{2}]\subset(0,\,T)$, in particular $T_0:=0$ and $T_{N+1}=T$. Here $\delta$ represents the time scale at which we have variations of the permittivity. We are interested in regimes where 
			\begin{equation*}
				\delta \ll  T.
			\end{equation*}
			
			\item Let $d:=\min_{0\leq i\leq N} d_i$, where $d_i:=T_{i+1}-T_i$, satisfy 
			\begin{equation*}\label{scale-d}
				d =O(\delta^l),\ l\in (0,\,1].
			\end{equation*}
		\end{enumerate}
	\end{defn}
	The analysis provided in this work is based on the large scales in (\ref{material-time-variable}). According to the modeling of $\omega_p$, see \cite{Kalluri-book}, this large scale is translated into the increase of the electron density.
	\bigskip
	
	We rewrite \eqref{integral-representation} in the following form
	\begin{equation}\label{system-1}
		\widetilde{E}(t) -C\delta^{-h} \sum^N_{j=1}\int^{T_j+\delta/2}_{T_j-\delta/2} \Phi(t,\,s)\widetilde{E}(s)ds=\widetilde{E}_{inc}(t),\quad t\in\mathbb{R}.
	\end{equation}


Our goal now is to analyze the asymptotic behavior of the total field $\widetilde{E}$, solution of (\ref{system-1}), in the regime described in Definition \ref{defn1}. The result is stated in the following theorem.
\begin{theorem}\label{results} Let the plasmonic frequency $\omega^2_p(t)$ be given as in (\ref{material-time-variable}) and assume the properties described in Definition \ref{defn1} be satisfied. Then, when $N$ is large enough, 
we have
\begin{equation}\label{result-equation}
	\widetilde{E}(t)
	=\widetilde{E}_{inc}(t) + C\delta^{1-h}\hat{F}\bar{A}^{-1}E^{in} +O\left(\delta^{\frac{3-h-l}{2}}\right),
\end{equation}
where $\hat{F}(t):=(\widetilde{\Phi}(t,\,T_1),\, \widetilde{\Phi}(t,\,T_2),\, \dots,\, \widetilde{\Phi}(t,\,T_N))$ with $\widetilde{\Phi}(t,\,s)=e^{i\kappa|t-s|}$, $\bar{A}^{-1}$ is the inverse of the matrix $\bar{A}$ with
\begin{eqnarray*}
	&&\bar{A}:=\left(\begin{array}{cccc}
		1-\beta & -\beta\tilde{\Phi}(T_1,\,T_2) & \cdots & -\beta\tilde{\Phi}(T_1,\,T_N)\\
		-\beta\tilde{\Phi}(T_2,\,T_1) & 1-\beta & \cdots & -\beta\tilde{\Phi}(T_2,\,T_N)\\
		\vdots & \vdots &\ddots & \vdots\\
		-\beta\tilde{\Phi}(T_N,\,T_1) & -\beta\tilde{\Phi}(T_N,\,T_2) & \cdots & 1-\beta\\
	\end{array}\right),
\end{eqnarray*}
$\beta:=\frac{iC\delta^{1-h}}{2\kappa}$ and $E^{in}:=(e^{i\kappa T_1},\,e^{i\kappa T_2}\,\dots,\,e^{i\kappa T_N})^T$.
\end{theorem}
The first interest in this result is that we can estimate the electric field after a succession (in time) of an arbitrary and sudden appearance (and then dis-appearance) of inhomogeneities. Its estimation is related to inverting an algebraic system, i.e. inverting the matrix $\bar{A}$. This provides a general form solution for the electromagnetism in the time-dependent permittivity, in the form of time-steps, compare with \cite{Kalluri-book, Caloz-Leger-2020} and the references therein.
\bigskip

\noindent A second interest in this result is the possibility to derive the effective permittivity as a substitute to multi-steps inhomogeneities when the number of step gets quite large and then characterize the generated equivalent electromagnetic field. 
\newline

\noindent Indeed, from \eqref{result-equation}, with $||\hat{F}||_2=O(N^{\frac{1}{2}})$, $||\bar{A}^{-1}||_2=O(1)$ and $||E^{in}||_2=O(N^{\frac{1}{2}})$, we have $||C\delta^{1-h}\hat{F}\bar{A}^{-1}E^{in}||_2=O\left(\delta^{(1-h-l)}\right)$. Therefore when $1-h-l>0$, we deduce that
\begin{equation*}
\widetilde{E}(t)\to\widetilde{E}_{inc}(t),\ \mbox{as}\ \delta\to0,\ \mbox{uniformly in terms of}\ t,
\end{equation*}
showing that there are almost no reflected waves.
Now, if $1-h-l\leq 0$, then the effective field  is different from the incident field. We characterize this effective field in the next theorem.
\begin{theorem}\label{effective-result}
For $N\in\mathbb{N}$, $N\gg1$, we have the following approximation 
\begin{equation*}
	\widetilde{E}(t)-\widetilde{E}_{eff}(t)=\begin{cases}
		O(\delta^{\frac{1-h+2l}{2}})+O(\delta^{1-h}),\quad &\mbox{if $1-h-l\geq0$,}\\
		O(\delta^{2-2h-l}), \quad &\mbox{if $\lambda T$ is close to $n\pi$, $1-h<l<2(1-h)$},\\
		O(\delta^{\frac{3-3h-l}{2}}), \quad &\mbox{if $\lambda T$ is away from $n\pi$, $1-h<l<3(1-h)$},
	\end{cases}
\end{equation*}
where $\widetilde{E}_{eff}$ satisfies 
\begin{equation*}
	\partial_t^2\widetilde{E}_{eff}(t)+\kappa^2\widetilde{E}_{eff}(t)+\left[C\delta^{-\alpha}\right]\chi_{[0,\,T]}\widetilde{E}_{eff}(t)=0, \quad t\in(-\infty,\,\infty),
\end{equation*}
with the exact solutions of the form
\begin{equation}\label{effective-field}
	\widetilde{E}_{eff}(t)=
	\begin{cases}
		\displaystyle e^{i\kappa t} +\frac{(\lambda^2-\kappa^2)(e^{i\lambda T}-e^{-i\lambda T})} {(\lambda+\kappa)^2e^{-i\lambda T}-(\lambda-\kappa)^2e^{i\lambda T}}e^{-i\kappa t},\quad & t\in(-\infty,\,0),\\
		\displaystyle \frac{2\kappa(\lambda+\kappa)e^{-i\lambda T}} {(\lambda+\kappa)^2e^{-i\lambda T}-(\lambda-\kappa)^2e^{i\lambda T}}e^{i\lambda t} +\frac{2\kappa(\lambda-\kappa)\,e^{i\lambda T}} {(\lambda+\kappa)^2e^{-i\lambda T}-(\lambda-\kappa)^2e^{i\lambda T}}e^{-i\lambda t},\quad & t\in[0,\,T],\\
		\displaystyle \frac{4\kappa\lambda e^{-i\kappa T}} {(\lambda+\kappa)^2e^{-i\lambda T}-(\lambda-\kappa)^2e^{i\lambda T}}e^{i\lambda t},\quad & t\in(T,\,\infty),
	\end{cases}
\end{equation}
where $\lambda:=\sqrt{\kappa^2+C\delta^{-\alpha}}$ and $\alpha:=-1+h+l$.
\end{theorem}
\bigskip

\noindent 
It is clear that, if $1-h-l>0$, i.e. $\alpha<0$, then $\lambda$ tends to $\kappa$, as $\delta \ll1 $, and hence the total field above tends to the incident field $e^{i \kappa t}$. When $1-h-l\leq 0$, i.e. $\alpha \geq 0$, then the dominant term of the effective field further simplifies as follows  
\begin{eqnarray*}
&&\widetilde{E}_{eff}(t)=
	\begin{cases}
		e^{i\kappa t}+o(1),\quad & t\in(-\infty,\,0),\\
		e^{i\lambda t}+o(1),\quad & t\in[0,\,T],\\
		\pm e^{i\kappa (t-T)}+o(1),\quad  & t\in(T,\,\infty),
	\end{cases}\quad\mbox{if $\lambda T$ is close to $n\pi$ and $1-h-l=0$,~ \footnotemark[3]} \\
&&\widetilde{E}_{eff}(t)=
	\begin{cases}
		e^{i\kappa t}+O(1)e^{-i\kappa t},\quad & t\in(-\infty,\,0),\\
		O(1)e^{i\lambda t}+O(1)e^{-i\lambda t},\quad & t\in[0,\,T],\\
		O(1)e^{i\kappa(t-T)}+o(1),\quad & t\in(T,\,\infty),
	\end{cases}\quad \mbox{if $\lambda T$ is away from $n\pi$ and $1-h-l=0$,}\\
&&\widetilde{E}_{eff}(t)=
\begin{cases}
	e^{i\kappa t}+o(1),\quad & t\in(-\infty,\,0),\\
	\frac{1}{2}e^{i\lambda t}+\frac{1}{2}e^{-i\lambda t}+O(\delta^{\frac{\alpha}{2}}),\quad & t\in[0,\,T],\\
	\pm e^{i\kappa (t-T)}+o(1),\quad  & t\in(T,\,\infty),
\end{cases}\quad\mbox{if $\lambda T$ is close to $n\pi$ and $1-h-l<0$,~ \footnotemark[3]} \\
&&\widetilde{E}_{eff}(t)=
\begin{cases}
	e^{i\kappa t}-e^{-i\kappa t}+O(\delta^{\frac{\alpha}{2}})e^{-i\kappa t},\quad & t\in(-\infty,\,0),\\
	O(\delta^{\frac{\alpha}{2}})e^{i\lambda t}+O(\delta^{\frac{\alpha}{2}})e^{-i\lambda t},\quad & t\in[0,\,T],\\
	O(\delta^{\frac{\alpha}{2}})e^{i\kappa(t-T)},\quad & t\in(T,\,\infty),
\end{cases}\quad \mbox{if $\lambda T$ is away from $n\pi$ and $1-h-l<0$,}
\end{eqnarray*}\footnotetext[3]{For instance, $\lambda T=m\pi+o(m^{-1})$.}where $n\in\mathbb{N}$, $n\gg1$ and $\pm$ for even or odd $n$. 

From the above form, we deduce that $\widetilde{E}_{eff}(t)$ is close to $\widetilde{E}_{inc}(t)$ for $1-h-l>0$, since in this case $\lambda\to\kappa$. If $1-h-l=0$, then it is related to a moderate, and not small, effective medium. In this case, both the reflected and the transmitted parts of the field are moderate, and not negligible, if $\lambda T$ is away from $n \pi$, for $n \in \mathbb{N}$,. If $\lambda T$ is close to $n \pi$, for $n \in \mathbb{N}$, then the wave field is {\it{fully transmitted}}.  
For $1-h-l<0$, however, it can be concluded that there are two different (and opposite) behaviors of effective field according to the values of $\lambda T$. To describe them let us recall that the incident wave is $\widetilde{E}_{inc}(t):=e^{i\kappa t}$, the reflected wave is proportional to $\widetilde{E}_{ref}(t):=e^{-i\kappa t}$ and the transmitted wave, in the part $(T, \infty)$, is proportional to $\widetilde{E}_{tra}(t):=e^{i\kappa t}$.
\begin{enumerate} 
\item When $\lambda T$ is close to the special values $n \pi$, $n \in \mathbb{N}$, recalling that $\lambda=\sqrt{\kappa^2+\omega_p^2}$, the incident wave is fully transmitted inside the effective medium without reflection. In this case, we have {\it{full transmission}}. 
In addition, we observe that the amplitude of the wave decreases and the oscillations increase within the effective medium, and both of them return to the same amplitude and frequency as the incident wave after passing through it.

\item When $\lambda T$ is away from $n \pi$, $n \in \mathbb{N}$, then the wave field $\widetilde{E}_{eff}$ is {\it{fully reflected}}. Indeed, in this case the transmitted part of the wave field is decreasing as $\delta\ll 1$.
\end{enumerate}
Therefore, we can modulate the plasmonic frequency in each time interval, so that the medium behaves as a {\it{wall}}, with full reflection, or as a {\it{well}}, i.e. with full transmission. This result is reminiscent to the known wall/well effects in quantum mechanics. The full transmission can be achieved only for a modulated permittivity close to some discrete values. These values can be understood as resonant frequencies. 

\subsection{A numerical validation}

\begin{center}
	\textbf{Table 1.} The parameter settings for three effective fields\\[+1mm]
	\begin{tabular}{ccccccccc}\hline\hline\\[-3.5mm]
		$T$ & $l$ & $h$ & $\delta$ & $\kappa$ & $C$ & $\lambda$ & $\lambda T$ & $\tan(\lambda T)$\\ \hline
		10 & 0.1 & 0.821 & 1e-03 & 1 & 1 & 1.2568 & 4.0004$\pi$ & 0.0012 \\
		
		10 & 0.1 & 0.5 & 1e-07 & 1 & 1 & 1.0008 & 3.1856$\pi$ & 0.6597 \\
		
		10 & 0.9 & 0.717 & 1e-03 & 1 & 1 & 8.4828 & 27.0016$\pi$ & 0.0049 \\
		
		10 & 0.9 & 0.369 & 1e-07 & 1 & 1 & 8.7968 & 28.0011$\pi$ & 0.0033 \\
		
		10 & 0.9 & 0.538 & 1e-07 & 1 & 1 & 34.1139 & 108.6517$\pi$ & -1.9368 \\ \hline\hline
	\end{tabular}
\end{center}

\begin{figure}[htp]
	\begin{center}
		\includegraphics[width=0.31\textwidth,height=4.2cm]{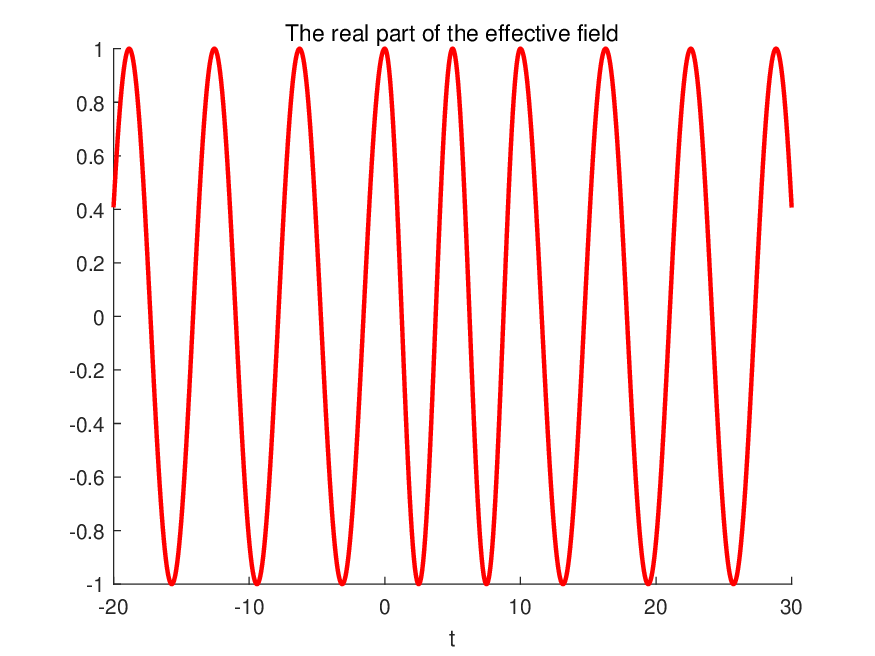}
		\includegraphics[width=0.31\textwidth,height=4.2cm]{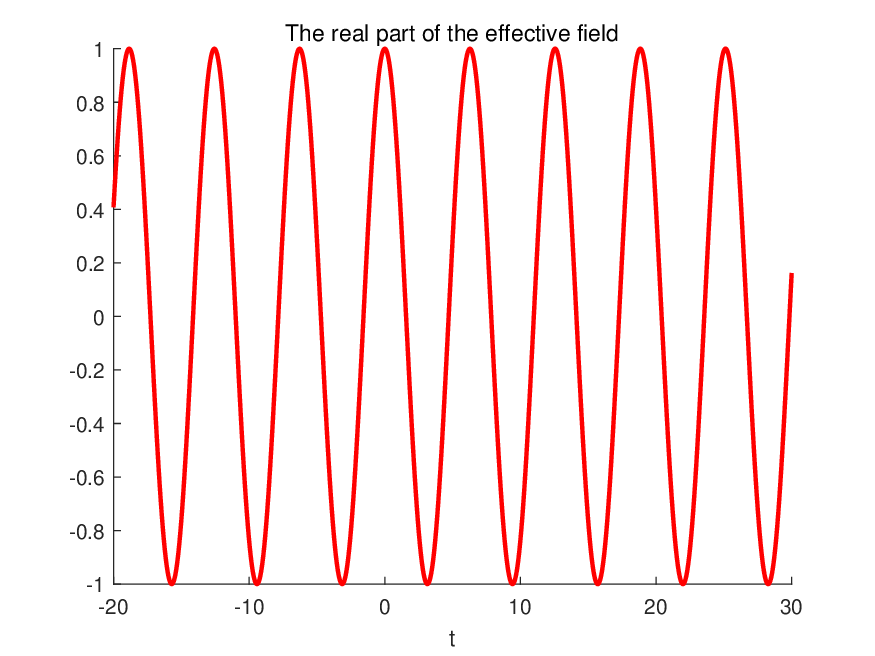}

		\includegraphics[width=0.31\textwidth,height=4.2cm]{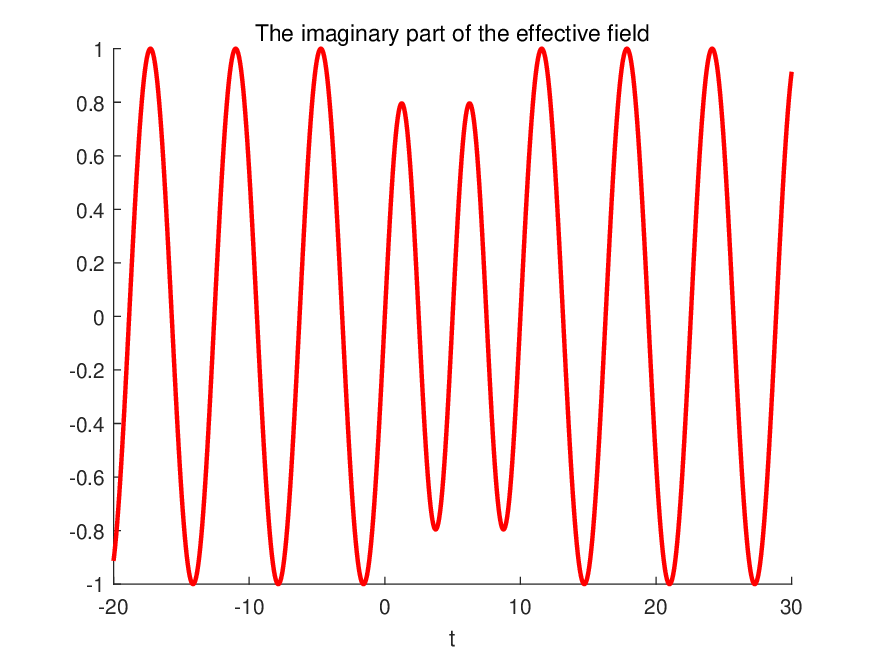}
		\includegraphics[width=0.31\textwidth,height=4.2cm]{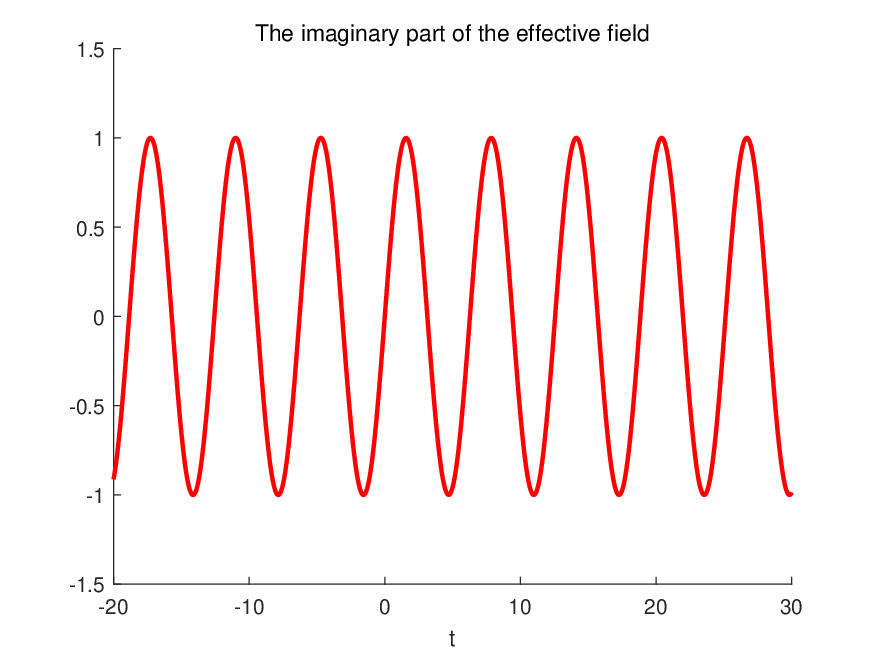}
		\caption{Numerical results for the effective field $\tilde{E}_{eff}$ when $1-h-l>0$: for $\lambda$ is close to even $n\pi$ (left), for $\lambda$ is away from $n\pi$ (right).}\label{exam0-p-1}
	\end{center}
\end{figure}

\begin{figure}[htp]
	\begin{center}
		\includegraphics[width=0.31\textwidth,height=4.2cm]{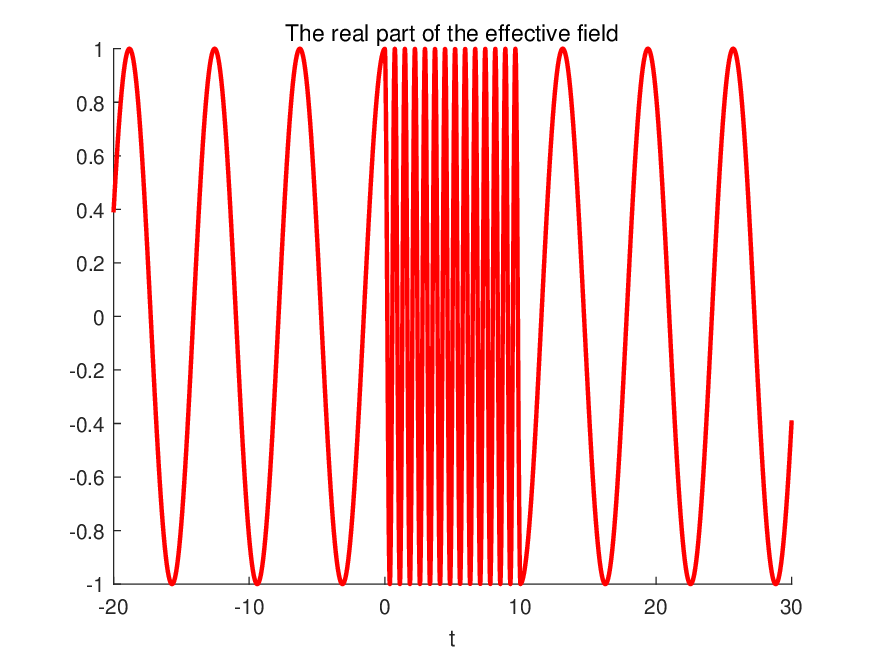}
		\includegraphics[width=0.31\textwidth,height=4.2cm]{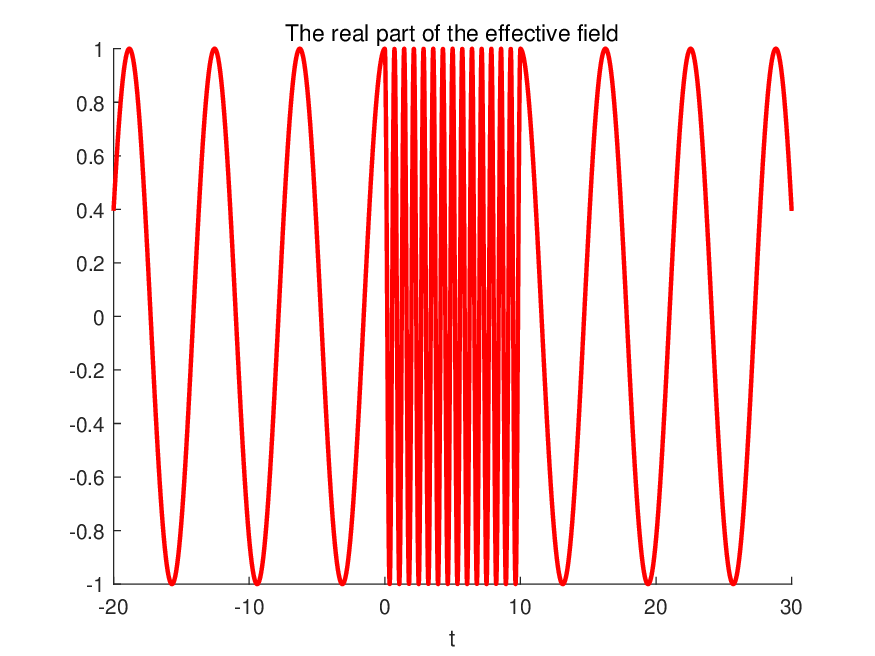}
		\includegraphics[width=0.31\textwidth,height=4.2cm]{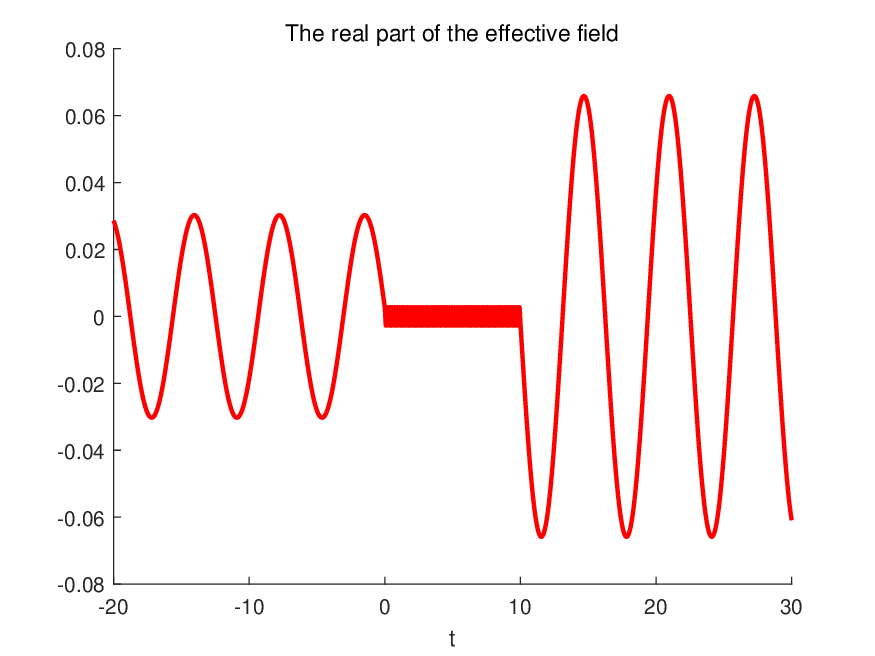}

		\includegraphics[width=0.31\textwidth,height=4.2cm]{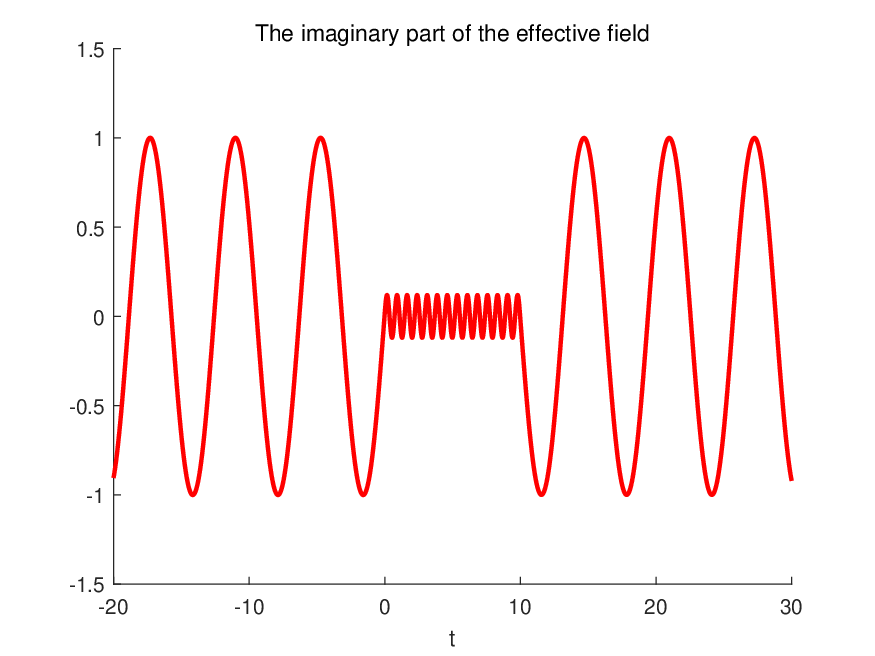}
		\includegraphics[width=0.31\textwidth,height=4.2cm]{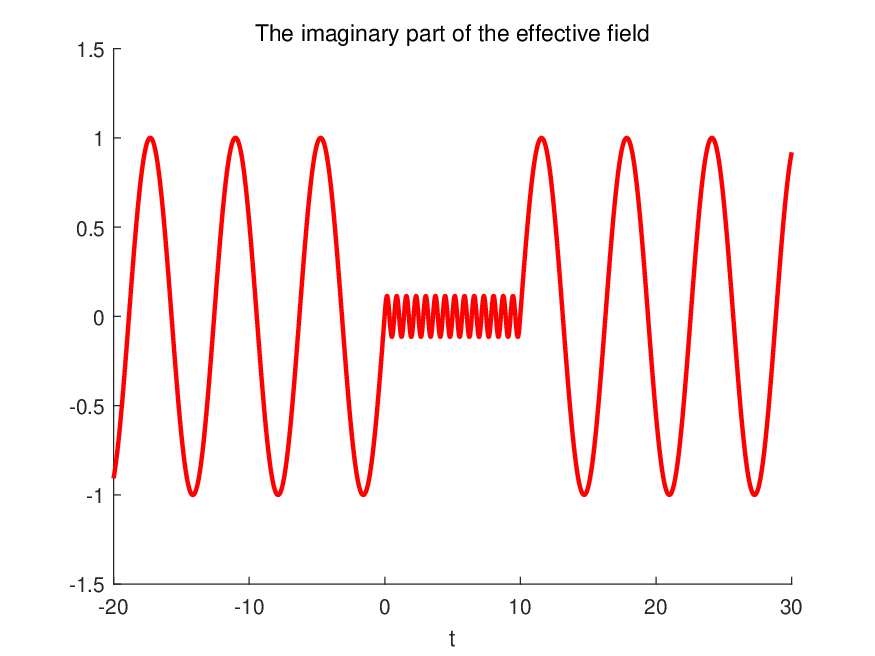}
		\includegraphics[width=0.31\textwidth,height=4.2cm]{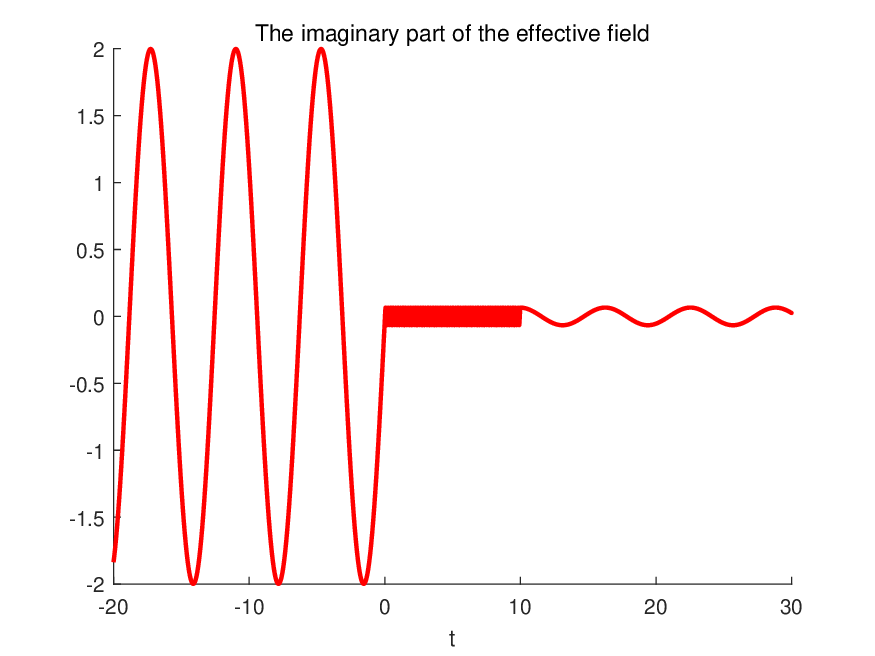}
		\caption{Numerical results for the effective field $\tilde{E}_{eff}$ when $1-h-l<0$: for $\lambda$ is close to odd $n\pi$ (left), for $\lambda$ is close to even $n\pi$ (middle), for $\lambda$ is away from $n\pi$ (right).}\label{exam0-p-2}
	\end{center}
\end{figure}

In Figure \ref{exam0-p-1}--\ref{exam0-p-2}, we provide few numerical tests supporting the results and discussions presented in Section \ref{Statement-results} focusing only on the behavior of the effective field. More numerical tests on the validity of the Foldy-Lax field approximation and its links to the effective field are provided in Section \ref{Numerical-tests}. 
\bigskip

Table 1 provides the parameter values for five cases under the assumptions described in Definition \ref{defn1}. Whether $\lambda T$ is close to $n\pi$ or not, it can be seen from Figure \ref{exam0-p-1} that the effective field is similar to the incident field. 
In the left an middle part of Figure \ref{exam0-p-2}, we see the full transmission for $1-h-l<0$ as $\lambda T$ is close to $n \pi$. Obviously, when $t<0$ and $t>T$, the wave is similar as the incident wave. But $\lambda T$ is close to $n \pi$ with different parity, which will produce the opposite directions at $t=T$. For the right part of Figure \ref{exam0-p-2}, as $\lambda T$ is away from $n \pi$, we observe a full reflection of the wave field.
\bigskip

The rest of the paper is organized as follows. In Section 2, we analyze the asymptotic behavior of the total field $\widetilde{E}$. In Section 3, we provide the analysis of the effective field as the proof of Theorem \ref{effective-result}. Several numerical examples are presented in Section 4 to illustrate the consistency of the asymptotic expansion and the effective field.

\section{Proof of Theorem \ref{results}}

The proof of Theorem \ref{results} is divided into two steps. In the first one we derive the needed a priori estimation of $\sum_{j=1}^N||\widetilde{E}||_{L^2(T_j-\delta/2,\,T_j+\delta/2)}$. To do this, we need to state and inverse a related linear algebraic system. Based on this estimate, in the second step, we complete the proof of Theorem \ref{results}.

\subsection{Step 1. The existence and uniqueness of solution of \eqref{integral-representation} }
\begin{prop}
	The solution of integral equation \eqref{integral-representation} exists and is unique from $L^2(0,\,T)$ to $L^2(0,\,T)$.
\end{prop}
{\bf Proof.}
	We rewrite \eqref{integral-representation} as
\begin{equation}\label{G1}
	\widetilde{E}(t) -C\delta^{-h} \int^{T}_{0} \Phi(t,\,s)\widetilde{E}(s)ds=\widetilde{E}_{inc}(t),\quad t\in\mathbb{R}.
\end{equation}
We assume $\widetilde{E}_{inc}\in C^2(0,\,T)$. Applying in \eqref{G1} the operator $\partial^2_t+\kappa^2$, we get 
\begin{equation}\label{G2}
	\partial^2_t\widetilde{E}(t)+\kappa^2\widetilde{E}(t)+C\delta^{-h}\widetilde{E}(t)ds=\partial^2_t\widetilde{E}_{inc}(t)+\kappa^2\widetilde{E}_{inc}(t), \quad t\in(0,\,T).
\end{equation}
Define $\bar{\lambda}:=\sqrt{\kappa^2+C\delta^{-h}}$ and $f(t):=\partial^2_t\widetilde{E}_{inc}(t)+\kappa^2\widetilde{E}_{inc}(t)$. It is clear that 
$$\widetilde{E}_p(t):=\int_0^T -\frac{i}{2\bar{\lambda}}e^{i\bar{\lambda}|t-s|}f(s)ds$$ is a solution of \eqref{G2}.
Observe that 
\begin{eqnarray*}
	\widetilde{E}_p(t)&=&-\frac{i}{2\bar{\lambda}}\int_0^t e^{i\bar{\lambda}(t-s)}f(s)ds -\frac{i}{2\bar{\lambda}}\int_t^Te^{-i\bar{\lambda}(t-s)}f(s)ds\\
	&=&-\frac{i}{2\bar{\lambda}}e^{i\bar{\lambda} t}\int_0^t e^{-i\bar{\lambda} s}f(s)ds -\frac{i}{2\bar{\lambda}}e^{-i\bar{\lambda} t}\int_t^Te^{i\bar{\lambda} s}f(s)ds.
\end{eqnarray*}
In addition, 
\begin{eqnarray*}
	\int_0^t e^{-i\bar{\lambda} s}\partial^2_s\widetilde{E}_{inc}(s)ds &=&e^{-i\bar{\lambda} t}\partial_t\widetilde{E}_{inc}(t)+i\bar{\lambda}\int_0^te^{-i\bar{\lambda} s}\partial_s\widetilde{E}_{inc}(s)ds\\
	&=&e^{-i\bar{\lambda} t}\partial_t\widetilde{E}_{inc}(t)+i\bar{\lambda} e^{-i\bar{\lambda} t}\widetilde{E}_{inc}(t) -\bar{\lambda}^2\int_0^te^{-i\bar{\lambda} s}\widetilde{E}_{inc}(s)ds,
\end{eqnarray*}
and 
\begin{eqnarray*}
	\int_t^T e^{i\bar{\lambda} s}\partial^2_s\widetilde{E}_{inc}(s)ds &=&-e^{i\bar{\lambda} t}\partial_t\widetilde{E}_{inc}(t)-i\bar{\lambda}\int_t^Te^{i\bar{\lambda} s}\partial_s\widetilde{E}_{inc}(s)ds\\
	&=&-e^{i\bar{\lambda} t}\partial_t\widetilde{E}_{inc}(t)+i\bar{\lambda} e^{i\bar{\lambda} t}\widetilde{E}_{inc}(t) -\bar{\lambda}^2\int_t^Te^{i\bar{\lambda} s}\widetilde{E}_{inc}(s)ds.
\end{eqnarray*}
Therefore,
\begin{equation}\label{G3}
	\widetilde{E}_p(t)= \widetilde{E}_{inc}(t)+(\bar{\lambda}^2-\kappa^2)\int_0^T \frac{i}{2\bar{\lambda}}e^{i\bar{\lambda}|t-s|}\widetilde{E}_{inc}(s)ds.
\end{equation}
Then, using the periodicity of $e^{is}$, we show that 
\begin{equation}\label{G4}
	||\widetilde{E}_p||_{L^2(0,\,T)}=O(T)||\widetilde{E}_{inc}||_{L^2(0,\,T)}.
\end{equation}
Hence, by the density of $C^2(0,\,t)$ in $L^2(0,\,t)$, we deduce that $\widetilde{E}_p(t)$ is also a solution of \eqref{G1} even if $\widetilde{E}_{inc}\in L^2(0,\,T)$.

Take $\widetilde{E}_{inc}\equiv0$ in \eqref{G1}. Then, we have \eqref{G2} as 
\begin{equation*}
	\partial^2_tE(t)+\bar{\lambda}^2E(t)=0.
\end{equation*}
Therefore, the solution is 
\begin{equation*}
	E(t)=C_1e^{i\bar{\lambda} t}+C_2e^{-i\bar{\lambda} t},
\end{equation*}
where $C_1$ and $C_2$ are constants.
However, from \eqref{G1}, we see that $E$ extends to $t<0$ as a function of the form $e^{i\bar{\lambda} t}$. Therefore, $C_2\equiv0$.
From \eqref{G1},  that $E$ extends to $t>T$ as a function of the form $e^{-i\bar{\lambda} t}$. Therefore, $C_1\equiv0$. Hence, we have $E\equiv0$.

Finally, the unique solution of \eqref{G1} is given by \eqref{G3}.\hfill $\Box$

\subsection{Step 2. The a priori estimate and the related linear algebraic system}
We start from the system \eqref{system-1} for $t\in I_m$, $m=1,\dots,N$,
\begin{equation}\label{system-1-N}
\widetilde{E}(t) -C\delta^{-h}\sum^N_{j=1} \int^{T_j+\delta/2}_{T_j-\delta/2} \Phi(t,\,s)\widetilde{E}(s)ds =\widetilde{E}_{inc}(t).
\end{equation}
For $t\in I_m$, $m=1,\dots,N$, define $\bar{E}_m(\bar{t}):=\widetilde{E}(t)$ and $\bar{E}^m_{inc}(\bar{t}):=\widetilde{E}_{inc}(t)$ with $\bar{t} :=\delta^{-1}(t-T_m)\in(-\frac{1}{2},\,\frac{1}{2})$. For $s\in I_j$, $j=1,\dots,N$, set $\bar{s} :=\delta^{-1}(s-T_j)\in(-\frac{1}{2},\,\frac{1}{2})$. Then, for $t\in I_m$ and $s\in I_j$, $m,j=1,\dots,N$, we denote
\begin{equation*}
\Phi(t,\,s) =\frac{i}{2\kappa}e^{i\kappa|t-s|} =\frac{i}{2\kappa}e^{i\kappa|\delta(\bar{t}-\bar{s})+T_m-T_j|} =:\bar{\Phi}_{m,j}(\bar{t},\,\bar{s}).
\end{equation*}
We rewrite \eqref{system-1-N} by the new denotation as
\begin{equation}\label{system-bar-N}
	\bar{E}_m(\bar{t}) =\bar{E}^m_{inc}(\bar{t}) +C\delta^{1-h} \sum_{j=1}^N \int^{\frac{1}{2}}_{-\frac{1}{2}} \bar{\Phi}_{m,j}(\bar{t},\,\bar{s})\bar{E}_j(\bar{s})d\bar{s}.
\end{equation}
As the expansion of $\bar{\Phi}_{m,j}(\bar{t},\,\bar{s})$ about $\bar{s}$ at the point $\bar{t}$
\begin{equation*}
\bar{\Phi}_{m,j}(\bar{t},\,\bar{s}) =\sum_{l=0}^\infty \frac{\partial^l_{\bar{s}}\bar{\Phi}_{m,j}(\bar{t},\,\bar{t})}{l!} (\bar{s}-\bar{t})^l =\frac{i}{2\kappa}\widetilde{\Phi}(T_m,\,T_j)  +\frac{i}{2\kappa}\widetilde{\Phi}(T_m,\,T_j)  \sum_{l=1}^\infty \frac{(i\kappa\delta)^l}{l!}\left(\frac{-T_m+T_j}{|T_m-T_j|}\right)^l(\bar{s}-\bar{t})^l,
\end{equation*}
where 
denote $\widetilde{\Phi}(t,\,s):=e^{i\kappa|t-s|}$, i.e., $\Phi(t,\,s)=\frac{i}{2\kappa}\widetilde{\Phi}(t,\,s)$,
then, \eqref{system-bar-N} are rewritten as
\begin{eqnarray*}
\bar{E}_m(\bar{t}) =\bar{E}^m_{inc}(\bar{t}) &+&\frac{iC\delta^{1-h}}{2\kappa}
\sum_{j=1}^N \widetilde{\Phi}(T_m,\,T_j) \int^{\frac{1}{2}}_{-\frac{1}{2}} \bar{E}_j(\bar{t})d\bar{t}\\
&+&\frac{iC\delta^{1-h}}{2\kappa} \sum_{j=1}^N \widetilde{\Phi}(T_m,\,T_j) \int^{\frac{1}{2}}_{-\frac{1}{2}} \sum_{l=1}^\infty \frac{(i\kappa\delta)^l}{l!}\left(\frac{-T_m+T_j}{|T_m-T_j|}\right)^l(\bar{s}-\bar{t})^l \bar{E}_j(\bar{s})d\bar{s}.
\end{eqnarray*}
With the estimate
\begin{eqnarray}\label{estimate-int-bar-sum}
\Bigg|\int^{\frac{1}{2}}_{-\frac{1}{2}}\sum_{l=1}^\infty \frac{(i\kappa\delta)^l}{l!}\left(\frac{-T_m+T_j}{|T_m-T_j|}\right)^l(\bar{s}-\bar{t})^l \bar{E}_j(\bar{s})d\bar{s}\Bigg|
&\leq&\left(\int^{\frac{1}{2}}_{-\frac{1}{2}} \Bigg|\sum_{l=1}^\infty \frac{(i\kappa\delta)^l}{l!}|\bar{s}-\bar{t}|^l\Bigg|^2 d\bar{s}\right)^{\frac{1}{2}} \left(\int^{\frac{1}{2}}_{-\frac{1}{2}}|\bar{E}_j(\bar{s})|^2 d\bar{s}\right)^{\frac{1}{2}}\nonumber\\
&\leq&\sum_{l=1}^\infty\Bigg|\frac{(i\kappa\delta)^l}{l!}\Bigg|\, ||\bar{E}_j||_{L^2(-\frac{1}{2},\frac{1}{2})}=2\sum_{l=1}^\infty\Bigg|\frac{(i\kappa\delta)^l}{2l!}\Bigg|\, ||\bar{E}_j||_{L^2(-\frac{1}{2},\frac{1}{2})}\nonumber\\
&\leq&2\sum_{l=1}^\infty \Bigg|\left(\frac{i\kappa\delta}{2}\right)^l\Bigg|\, ||\bar{E}_j||_{L^2(-\frac{1}{2},\frac{1}{2})}\nonumber\\
&=&O(\delta)\,||\bar{E}_j||_{L^2(-\frac{1}{2},\frac{1}{2})},\quad \mbox{if}\ \delta<\frac{1}{\kappa},
\end{eqnarray}
we get 
\begin{eqnarray}\label{system-bar-withO}
\bar{E}_m(\bar{t}) =\bar{E}^m_{inc}(\bar{t}) &+&\frac{iC\delta^{1-h}}{2\kappa} \sum_{j=1}^N \widetilde{\Phi}(T_m,\,T_j) \int^{\frac{1}{2}}_{-\frac{1}{2}} \bar{E}_j(\bar{t})d\bar{t}\nonumber\\
&+&O\left(\delta^{2-h}\right) \sum_{j=1}^N \widetilde{\Phi}(T_m,\,T_j) ||\bar{E}_j||_{L^2(-\frac{1}{2},\frac{1}{2})}.
\end{eqnarray}
Then, integrate \eqref{system-bar-withO} on $\bar{t}$ from $-\frac{1}{2}$ to $\frac{1}{2}$ as
\begin{eqnarray}\label{a-system-1}
\int^{\frac{1}{2}}_{-\frac{1}{2}} \bar{E}_m(\bar{t})d\bar{t} =\int^{\frac{1}{2}}_{-\frac{1}{2}} \bar{E}^m_{inc}(\bar{t})d\bar{t} &+&\frac{iC\delta^{1-h}}{2\kappa} \sum_{j=1}^N \widetilde{\Phi}(T_m,\,T_j) \int^{\frac{1}{2}}_{-\frac{1}{2}} \bar{E}_j(\bar{t})d\bar{t}\nonumber\\
&+&O\left(\delta^{2-h}\right) \sum_{j=1}^N \widetilde{\Phi}(T_m,\,T_j) ||\bar{E}_j||_{L^2(-\frac{1}{2},\frac{1}{2})}.
\end{eqnarray}

Let $$\bar{q}_m :=\int^{\frac{1}{2}}_{-\frac{1}{2}} \bar{E}_m(\bar{t})d\bar{t},\quad \bar{p}_m :=||\bar{E}_m||_{L^2(-\frac{1}{2},\frac{1}{2})},\quad \bar{d}_m :=\int^{\frac{1}{2}}_{-\frac{1}{2}} \bar{E}^m_{inc}(\bar{t})d\bar{t}.$$
Then we derive the algebraic system
\begin{equation}\label{algebraic-system}
	\bar{A}\bar{q} =O\left(\delta^{2-h}\right)Q\bar{p} +\bar{d},
\end{equation}
where
\begin{eqnarray*}
	&&\bar{A}:=\left(\begin{array}{cccc}
		1-\beta & -\beta\tilde{\Phi}(T_1,\,T_2) & \cdots & -\beta\tilde{\Phi}(T_1,\,T_N)\\
		-\beta\tilde{\Phi}(T_2,\,T_1) & 1-\beta & \cdots & -\beta\tilde{\Phi}(T_2,\,T_N)\\
		\vdots & \vdots &\ddots & \vdots\\
		-\beta\tilde{\Phi}(T_N,\,T_1) & -\beta\tilde{\Phi}(T_N,\,T_2) & \cdots & 1-\beta\\
	\end{array}\right), \quad
	\bar{q}:=\left(\begin{array}{c}
		\bar{q}_1 \\
		\bar{q}_2 \\
		\vdots\\
		\bar{q}_N \\
	\end{array}\right), \quad
	\bar{d}:=\left(\begin{array}{c}
		\bar{d}_1 \\
		\bar{d}_2 \\
		\vdots\\
		\bar{d}_N \\
	\end{array}\right),\\
	&&Q:=\left(\begin{array}{cccc}
		1 & \tilde{\Phi}(T_1,\,T_2) & \cdots & \tilde{\Phi}(T_1,\,T_N)\\
		\tilde{\Phi}(T_2,\,T_1) & 1 & \cdots & \tilde{\Phi}(T_2,\,T_N)\\
		\vdots & \vdots &\ddots & \vdots\\
		\tilde{\Phi}(T_N,\,T_1) & \tilde{\Phi}(T_N,\,T_2) & \cdots & 1\\
	\end{array}\right), \quad
	\bar{p}:=\left(\begin{array}{c}
		\bar{p}_1 \\
		\bar{p}_2 \\
		\vdots\\
		\bar{p}_N \\
	\end{array}\right),\quad \beta:=\frac{iC\delta^{1-h}}{2\kappa}.
\end{eqnarray*}

\begin{prop}
	The matrix $\bar{A}$ is invertible and the $2$-norm satisfies the estimate $||\bar{A}^{-1}||_2=O(1)$.
\end{prop}
{\bf Proof.}
Rewrite \eqref{a-system-1} as
\begin{equation}\label{G5}
	\bar{q}_m-\bar{C}\delta^{-\alpha}\sum_{j=1}^N\widetilde{\Phi}(T_m,\,T_j)\delta^{l}\bar{q}_j=F_m,\quad m=1,\,\dots,\,N,
\end{equation}
where $\bar{C}:=\frac{iC}{2\kappa}$, $\alpha=-1+h+l$ and $$F_m:=\int^{\frac{1}{2}}_{-\frac{1}{2}} \bar{E}^m_{inc}(\bar{t})d\bar{t}+O\left(\delta^{2-h}\right) \sum_{j=1}^N \widetilde{\Phi}(T_m,\,T_j) ||\bar{E}_j||_{L^2(-\frac{1}{2},\frac{1}{2})}.$$
Then,
\begin{equation}\label{GG1}
\bar{q}_m-\bar{C}\delta^{-\alpha}\int_0^T\sum_{j=1}^N\widetilde{\Phi}(T_m,\,T_j)x_j\bar{q}_jds=F_m,\quad m=1,\,\dots,\,N,
\end{equation}
where $x_j:=\chi_{\bar{I}_j}$, $\chi$ is characteristic function and $\bar{I}_j:=(t_j,\,t_{j+1})$.
Rewrite \eqref{GG1} as
\begin{equation}\label{GG2}
	\bar{q}_m-\bar{C}\delta^{-\alpha}\int_0^T\widetilde{\Phi}(T_m,\,s)\sum_{j=1}^Nx_j\bar{q}_jds=F_m-\bar{C}\delta^{-\alpha}\int_0^T\sum_{j=1}^N\left(\widetilde{\Phi}(T_m,\,s)-\widetilde{\Phi}(T_m,\,T_j)\right)x_j\bar{q}_jds.
\end{equation}
Sum \eqref{GG2} together from $m=1$ to $m=N$,
\begin{eqnarray}\label{GG3}
	&&\sum_{m=1}^Nx_m\bar{q}_m -\bar{C}\delta^{-\alpha}\int_0^T\sum_{m=1}^Nx_m\widetilde{\Phi}(T_m,\,s)\sum_{j=1}^Nx_j\bar{q}_jds\nonumber\\ 
	&&=\sum_{m=1}^Nx_mF_m -\bar{C}\delta^{-\alpha}\int_0^T\sum_{j=1}^N\sum_{m=1}^N\left[x_m\left(\widetilde{\Phi}(T_m,\,s)-\widetilde{\Phi}(T_m,\,T_j)\right)\right]x_jx_j\bar{q}_jds.
\end{eqnarray}
Denote $\sum_{m=1}^Nx_m\bar{q}_m$ by $q$ and $\sum_{m=1}^Nx_mF_m$ by $F$.
With 
\begin{eqnarray*}
&&\int_0^T\sum_{j=1}^N\sum_{m=1}^N\left[x_m\left(\widetilde{\Phi}(T_m,\,s)-\widetilde{\Phi}(T_m,\,T_j)\right)\right]x_jx_j\bar{q}_jds\\
&&\leq\left[\int_0^T\left(\sum_{j=1}^N\sum_{m=1}^N\left[x_m\left(\widetilde{\Phi}(T_m,\,s)-\widetilde{\Phi}(T_m,\,T_j)\right)\right]x_j\right)^2ds\right]^{\frac{1}{2}}\left[\int_0^T\left(\sum_{j=1}^Nx_j\bar{q}_j\right)^2ds\right]^{\frac{1}{2}}\\
&&\leq\left[\int_0^T\left(\sum_{m=1}^Nx_m\left[\sum_{j=1}^N\widetilde{\Phi}(T_m,\,s)x_j-\sum_{j=1}^N\widetilde{\Phi}(T_m,\,T_j)x_j\right]\right)^2ds\right]^{\frac{1}{2}}||q||_{L^2(0,\,T)}\\
&&=\sum_{m=1}^Nx_m\left[\int_0^T\left(\widetilde{\Phi}(T_m,\,s)-\sum_{j=1}^N\widetilde{\Phi}(T_m,\,T_j)x_j\right)^2ds\right]^{\frac{1}{2}}||q||_{L^2(0,\,T)}.
\end{eqnarray*}
Now, $\forall r>0$ and $\exists N_0(\delta,\,\,r)$, $\forall N>N_0(\delta,\,\,r)$, such that
\begin{equation}
	\sup_{t\in(0,\,T)}\left(\int_0^T\left[\widetilde{\Phi}(t,\,s)-\sum_{j=1}^N\widetilde{\Phi}(t,\,T_j)x_j\right]^2ds\right)^{\frac{1}{2}}\lesssim\delta^r,
\end{equation}
rewrite \eqref{GG3} as
\begin{equation}\label{GG4}
	q(t)-\bar{C}\delta^{-\alpha}\int_0^T\widetilde{\Phi}(t,\,s)q(s)ds =F(t)+O(\delta^{r-\alpha})||q||_{L^2(0,\,T)}.    
\end{equation}

Let $A: L^2(0,\,T)\to L^2(0,\,T)$ be the operator appearing in \eqref{G1}. According to \eqref{G4}, we have $||A^{-1}||=O(1)$. Therefore, \eqref{GG4} is invertible and 
\begin{equation}
	||q||_{L^2(0,\,T)}\lesssim ||F||_{L^2(0,\,T)}+O(\delta^{r-\alpha})||q||_{L^2(0,\,T)}.
\end{equation}
Choosing $r>\alpha$, we deduce that 
\begin{equation}
	||q||_{L^2(0,\,T)}\lesssim ||F||_{L^2(0,\,T)}.
\end{equation}
But 
\begin{equation}
	||q||^2_{L^2(0,\,T)}\sim \sum_{m=1}^N|\bar{q}_m|^2.
\end{equation}
Then, \eqref{algebraic-system} is invertible and 
\begin{equation}\label{GG5}
	\left(\sum_{m=1}^N|\bar{q}_m|^2\right)^{\frac{1}{2}} \lesssim\left(\sum_{m=1}^N|F_m|^2\right)^{\frac{1}{2}}.
\end{equation}
The proof is complete. \hfill $\Box$

\begin{prop}\label{estimate-norm-N}
For $N$ large enough, we have
\begin{equation}
	\sum^N_{j=1} ||\widetilde{E}||_{L^2(T_j-\delta/2,T_j+\delta/2)}=O(\delta^{\frac{1}{2}}N).
\end{equation}
\end{prop}
{\bf Proof.}
As the algebraic system \eqref{algebraic-system} is invertible, we have
\begin{equation}\label{matrix-q-p}
\bar{q}=O\left(\delta^{2-h}\right)\bar{A}^{-1}Q\bar{p} +\bar{A}^{-1}\bar{d}.
\end{equation}
Now from \eqref{system-bar-withO}, we get
\begin{eqnarray*}
||\bar{E}_m||_{L^2(-\frac{1}{2},\frac{1}{2})} \leq ||\bar{E}^m_{inc}||_{L^2(-\frac{1}{2},\frac{1}{2})} &+&O\left(\delta^{1-h}\right)\sum_{j=1}^N |\widetilde{\Phi}(T_m,\,T_j)|  \bigg|\int^{\frac{1}{2}}_{-\frac{1}{2}} \bar{E}_j(\bar{t})d\bar{t}\,\bigg|\nonumber\\
&+&O\left(\delta^{2-h}\right) \sum_{j=1}^N |\widetilde{\Phi}(T_m,\,T_j)|\, ||\bar{E}_j||_{L^2(-\frac{1}{2},\frac{1}{2})}.
\end{eqnarray*}
Let $\bar{b}_m :=||\bar{E}^m_{inc}||_{L^2(-\frac{1}{2},\frac{1}{2})}$.
Then we have 
\begin{equation*}
\bar{p}\leq\bar{b} +O\left(\delta^{1-h}\right)\mbox{abs}(Q)\,\mbox{abs}(\bar{q}) +O\left(\delta^{2-h}\right) \mbox{abs}(Q)\bar{p},
\end{equation*}
with $\bar{b}:=(\bar{b}_1,\,\bar{b}_2,\,\dots,\,\bar{b}_N)^T$, where $\mbox{abs}(X)$ denotes a matrix whose elements are the absolute values of the corresponding elements in $X$.
Based on \eqref{matrix-q-p} and each element in abs$(\bar{d})$ is smaller than $\bar{b}$, we get
\begin{eqnarray*}
\bar{p} &\leq&\bar{b}+O\left(\delta^{1-h}\right) \mbox{abs}(Q)\,\mbox{abs}\left(O\left(\delta^{2-h}\right)\bar{A}^{-1}Q\bar{p} +\bar{A}^{-1}\bar{d}\right) +O\left(\delta^{2-h}\right) \mbox{abs}(Q)\bar{p}\\
&\leq&\bar{b} +O(\delta^{3-2h}) \mbox{abs}(Q)\mbox{abs}(\bar{A}^{-1}Q)\bar{p} +O\left(\delta^{1-h}\right)\mbox{abs}(Q)\mbox{abs}(\bar{A}^{-1})\mbox{abs}(\bar{d}) +O\left(\delta^{2-h}\right) \mbox{abs}(Q)\bar{p}\\
&\leq&(I+O\left(\delta^{1-h}\right)\mbox{abs}(Q)\mbox{abs}(\bar{A}^{-1}))\bar{b} + \mbox{abs}(Q)\left[O\left(\delta^{3-2h}\right)\mbox{abs}(\bar{A}^{-1}Q)+O\left(\delta^{2-h}\right)I\right]\,\bar{p}.
\end{eqnarray*}
Then rewrite as
\begin{equation*}
\left[I-\mbox{abs}(Q)\left[O\left(\delta^{3-2h}\right)\mbox{abs}(\bar{A}^{-1}Q)+O\left(\delta^{2-h}\right)I\right]\right]\bar{p} \leq(I+O\left(\delta^{1-h}\right)\mbox{abs}(Q)\mbox{abs}(\bar{A}^{-1}))\bar{b}.
\end{equation*}
With $||Q||_2=O(1)$ and $||\bar{A}^{-1}||_2=O(1)$, we have
$\mbox{abs}(Q)\left[O\left(\delta^{3-2h}\right)\mbox{abs}(\bar{A}^{-1}Q)+O\left(\delta^{2-h}\right)I\right]$ is enough small,
and
\begin{equation*}
||I+O\left(\delta^{1-h}\right)\mbox{abs}(Q)\mbox{abs}(\bar{A}^{-1})||_2=O(1).
\end{equation*}
Hence, 
\begin{equation*}\label{est-p-eq}
||\bar{p}||_2 \leq C^{\prime\prime}||\bar{b}||_2,
\end{equation*}
where $C^{\prime\prime}$ is a constant. Finally, we have
\begin{eqnarray*}\label{estimate-norm-D-N>1-result}
\sum^N_{j=1} ||\widetilde{E}||_{L^2(T_j-\delta/2,T_j+\delta/2)}
&=&\delta^{\frac{1}{2}}\sum^N_{j=1} ||\bar{E}_j||_{L^2(-\frac{1}{2},\frac{1}{2})} \leq\delta^{\frac{1}{2}}N^{\frac{1}{2}}||\bar{p}||_2=O(\delta^{\frac{1}{2}}N^{\frac{1}{2}})||\bar{b}||_2\nonumber\\
&=&O(\delta^{\frac{1}{2}}N^{\frac{1}{2}})\left(\sum^N_{j=1} ||\bar{E}^j_{inc}||^2_{L^2(-\frac{1}{2},\frac{1}{2})}\right)^{\frac{1}{2}} =O(N^{\frac{1}{2}})\sum^N_{j=1} ||\widetilde{E}_{inc}||_{L^2(T_j-\delta/2,T_j+\delta/2)}\nonumber\\
&=&O(\delta^{\frac{1}{2}}N).
\end{eqnarray*}
The proof is complete. \hfill $\Box$

\subsection{Step 3. Completing the proof of Theorem \ref{results}}

For \eqref{system-1}, $t\in I_m$, $m=1,\dots,N$, it can be rewritten as
\begin{equation}\label{system-2-N}
\widetilde{E}(t) =\widetilde{E}_{inc}(t) +C\delta^{-h} \sum^N_{j=1}\int^{T_j+\delta/2}_{T_j-\delta/2} \Phi(t,\,T_j)\widetilde{E}(s)ds +C\delta^{-h} \sum^N_{j=1}\int^{T_j+\delta/2}_{T_j-\delta/2} [\Phi(t,\,s)-\Phi(t,\,T_j)]\widetilde{E}(s)ds.
\end{equation}
Using the Taylor expansion to $\Phi(t,\,s)$ about $s$ at the point $T_j$, $j=1,\dots,N$, we have
\begin{equation*}
\Phi(t,\,s)-\Phi(t,\,T_j)=(s-T_j)R_j(t,\,s),
\end{equation*}
where $R_j(t,s):=\int_0^1\partial_s \Phi(t,\,s-\alpha(s-T_j))\,d\alpha$,
and $|R_j(t,s)|\leq\frac{1}{2}$. 
With the Cauchy-Schwarz inequality, we have
\begin{eqnarray}\label{integral-phi-D_j}
\Big|\int^{T_j+\delta/2}_{T_j-\delta/2} [\Phi(t,\,s)-\Phi(t,\,T_j)]\widetilde{E}(s)ds\Big| &=&\Big|\int^{T_j+\delta/2}_{T_j-\delta/2} (s-T_j)R_j(t,s)\widetilde{E}(s)ds\Big|\nonumber\\
&\leq&\left(\int^{T_j+\delta/2}_{T_j-\delta/2} |s-T_j|^2|R_j(t,s)|^2ds\right)^{\frac{1}{2}} \left(\int^{T_j+\delta/2}_{T_j-\delta/2} |\widetilde{E}(s)|^2ds\right)^{\frac{1}{2}}\nonumber\\
&\leq&\left(\int^{T_j+\delta/2}_{T_j-\delta/2} \frac{\delta^2}{4}\frac{1}{4}ds\right)^{\frac{1}{2}} \left(\int^{T_j+\delta/2}_{T_j-\delta/2} |\widetilde{E}(s)|^2ds\right)^{\frac{1}{2}}\nonumber\\
&\leq&\frac{1}{4}\delta^{\frac{3}{2}}\, ||\widetilde{E}||_{L^2(T_j-\delta/2,T_j+\delta/2)}.
\end{eqnarray}
With the estimate of $||\widetilde{E}||_{L^2(T_j-\delta/2,T_j+\delta/2)}$ in Proposition \ref{estimate-norm-N}, \eqref{system-2-N} becomes 
\begin{equation}\label{caseN-system-1}
\widetilde{E}(t) = \widetilde{E}_{inc}(t) +C\delta^{-h}\sum^N_{j=1}\Phi(t,\,T_j) \int^{T_j+\delta/2}_{T_j-\delta/2}\widetilde{E}(s)ds +O\left(\delta^{2-h}N\right), \quad t\in I_m,\ m=1,\dots,N.
\end{equation}
Integrate \eqref{caseN-system-1} about $t$ from $T_m-\delta/2$ to $T_m+\delta/2$ for $m=1,\dots,N$ respectively as
\begin{equation*}
	\int^{T_m+\delta/2}_{T_m-\delta/2}\widetilde{E}(t)dt =\int^{T_m+\delta/2}_{T_m-\delta/2}\widetilde{E}_{inc}(t)dt +C\delta^{-h} \sum_{j=1}^N\int^{T_m+\delta/2}_{T_m-\delta/2}\Phi(t,\,T_j)dt \int^{T_j+\delta/2}_{T_j-\delta/2}\widetilde{E}(s)ds 
	+O\left(\delta^{3-h}N\right).
\end{equation*}
Taking the Taylor series of $\Phi(t,\,T_j)$, $j=1,\dots,N$ about $t$ at the point $T_m$, $m=1,\dots,N$, as
\begin{equation*}
\Phi(t,\,T_j)=\sum_{l=0}^{\infty} \frac{\partial_t^l\Phi(T_m,\,T_j)}{l!}(t-T_m)^l =\Phi(T_m,\,T_j)\left(1+\sum_{l=1}^{\infty} \frac{(i\kappa)^l}{l!}|t-T_m|^l\right),
\end{equation*}
we have
\begin{eqnarray}\label{try-multiple-12}
\int^{T_m+\delta/2}_{T_m-\delta/2}\widetilde{E}(t)dt &=&\int^{T_m+\delta/2}_{T_m-\delta/2}\widetilde{E}_{inc}(t)dt +\frac{iC\delta^{1-h}}{2\kappa}\sum_{j=1}^N\widetilde{\Phi}(T_m,\,T_j)\int^{T_j+\delta/2}_{T_j-\delta/2}\widetilde{E}(s)ds +O\left(\delta^{3-h}N\right)\nonumber\\
&&+\frac{iC\delta^{-h}}{2\kappa} \sum_{j=1}^N \widetilde{\Phi}(T_m,\,T_j) \int^{T_m+\delta/2}_{T_m-\delta/2} \sum_{l=1}^{\infty}\frac{(i\kappa)^l}{l!}|t-T_m|^ldt \int^{T_j+\delta/2}_{T_j-\delta/2}\widetilde{E}(s)ds.
\end{eqnarray}
With 
\begin{equation*}
\bigg|\int^{T_j+\delta/2}_{T_j-\delta/2}\widetilde{E}(s)ds\bigg|\leq \delta^{\frac{1}{2}}  ||\widetilde{E}||_{L^2(T_j-\delta/2,T_j+\delta/2)},
\end{equation*}
and
\begin{eqnarray*}
\int^{T_m+\delta/2}_{T_m-\delta/2} \Big|\sum_{l=1}^{\infty} \frac{(i\kappa)^l}{l!}|t-T_m|^l\Big|dt
&\leq&\sum_{l=1}^{\infty}\frac{1}{l!}\left(\frac{\kappa\delta}{2}\right)^l \int^{T_m+\delta/2}_{T_m-\delta/2}dt\\
(2l!\geq2^l)&\leq&2\delta\sum_{l=1}^{\infty} \left(\frac{\kappa\delta}{4}\right)^l=O(\delta^2),\quad \mbox{if}\ \delta<\frac{1}{\kappa},
\end{eqnarray*}
\eqref{try-multiple-12} for $m=1,\dots,N$ can be simplified to
\begin{equation}\label{try-multiple-m}
\int^{T_m+\delta/2}_{T_m-\delta/2}\widetilde{E}(t)dt =\int^{T_m+\delta/2}_{T_m-\delta/2}\widetilde{E}_{inc}(t)dt +\frac{iC\delta^{1-h}}{2\kappa}  \sum_{j=1}^N\widetilde{\Phi}(T_m,\,T_j) \int^{T_j+\delta/2}_{T_j-\delta/2}\widetilde{E}(s)ds +O\left(\delta^{3-h}N\right).
\end{equation}
Let 
\begin{equation*}
\widetilde{q}_m :=\int^{T_m+\delta/2}_{T_m-\delta/2}\widetilde{E}(t)dt, \quad \widetilde{d}_m :=\int^{T_m+\delta/2}_{T_m-\delta/2}\widetilde{E}_{inc}(t)dt.
\end{equation*}
Based on the invertibility of $\bar{A}$, we have
\begin{equation}\label{matrix-q-d}
\widetilde{q}=\bar{A}^{-1}\widetilde{d} + O\left(\delta^{3-h}N\right) \bar{A}^{-1}\widetilde{e},
\end{equation}
with $\widetilde{q}:=(\widetilde{q}_1,\,\widetilde{q}_2,\,\dots,\,\widetilde{q}_N)^T$, $\widetilde{d}:=(\widetilde{d}_1,\,\widetilde{d}_2,\,\dots,\,\widetilde{d}_N)^T$ and $\widetilde{e}:=(1,\,1,\,\dots,\,1)^T$.
For \eqref{integral-representation}, $t\in\mathbb{R}$, with \eqref{integral-phi-D_j}, we get
\begin{eqnarray}\label{try-multiple-1-3}
	\widetilde{E}(t)-\widetilde{E}_{inc}(t) 
	&=&C\delta^{-h} \sum_{m=1}^N \Phi(t,\,T_m) \int^{T_m+\delta/2}_{T_m-\delta/2}\widetilde{E}(t)dt +C\delta^{-h}\sum_{m=1}^N \int^{T_m+\delta/2}_{T_m-\delta/2} (\Phi(t,\,s)-\Phi(t,\,T_m))\widetilde{E}(s)ds\nonumber\\
	&=&C\delta^{-h} \sum_{m=1}^N \Phi(t,\,T_m) \int^{T_m+\delta/2}_{T_m-\delta/2}\widetilde{E}(t)dt +O\left(\delta^{2-h}N\right).
\end{eqnarray}
Denote $\hat{F}(t):=(\widetilde{\Phi}(t,\,T_1),\, \widetilde{\Phi}(t,\,T_2),\, \dots,\, \widetilde{\Phi}(t,\,T_N))$ with $||\hat{F}||_2 =(\sum_{n=1}^N|\widetilde{\Phi}^2(t,\,T_n)|)^{\frac{1}{2}}\leq N^{\frac{1}{2}}$.
Coupling \eqref{matrix-q-d} and \eqref{try-multiple-1-3}, according to $||\bar{A}^{-1}||_2=O(1)$ and $||\widetilde{e}||_2=N^{\frac{1}{2}}$, we have
\begin{eqnarray}\label{try-multiple-final}
\widetilde{E}(t)
&=&\widetilde{E}_{inc}(t) +C\delta^{-h}\hat{F}\widetilde{q} +O\left(\delta^{2-h}N\right)\nonumber\\
&=&\widetilde{E}_{inc}(t) +C\delta^{-h}\hat{F} \left(\bar{A}^{-1}\widetilde{d} + O\left(\delta^{3-h}N\right) \bar{A}^{-1}\widetilde{e}\right) +O\left(\delta^{2-h}N\right)\nonumber\\
&=&\widetilde{E}_{inc}(t) +C\delta^{-h}\hat{F}\bar{A}^{-1}\widetilde{d} + O\left(\delta^{3-2h}N||\hat{F}||_2\,||\bar{A}^{-1}||_2\,||\widetilde{e}||_2\right) +O\left(\delta^{2-h}N\right)\nonumber\\
&=&\widetilde{E}_{inc}(t) +C\delta^{-h}\hat{F}\bar{A}^{-1}\widetilde{d} +O\left(\delta^{3-2h}N^2\right) +O(\delta^{2-h}N).
\end{eqnarray}
In addition, we have $\widetilde{d}=\delta E^{in}+O(\kappa\delta^2)\widetilde{e}$, where $E^{in}:=(e^{i\kappa T_1},\,e^{i\kappa T_2}\,\dots,\,e^{i\kappa T_N})^T$, since
\begin{equation*}
\widetilde{d}_m=\int^{T_m+\delta/2}_{T_m-\delta/2}\widetilde{E}_{inc}(t)dt= \int^{T_m+\delta/2}_{T_m-\delta/2}e^{i\kappa T_m}dt +e^{i\kappa T_m}\int^{T_m+\delta/2}_{T_m-\delta/2} \sum_{l=1}^{\infty} \frac{(i\kappa)^l}{l!}(t-T_m)^ldt =\delta e^{i\kappa T_m}+O(\delta^2).
\end{equation*}
We get the simple form that
\begin{equation*}
\widetilde{E}(t)
=\widetilde{E}_{inc}(t) + C\delta^{1-h}\hat{F}\bar{A}^{-1}E^{in} +O\left(\delta^{2-l-h}\right) +O\left(\delta^{3-2l-2h}\right).
\end{equation*}
The proof is complete. \hfill $\Box$

\section{The effective field}
In this section, we prove Theorem \ref{effective-result}. Consider the integral equation
\begin{equation}\label{eff-ori-system}
\widetilde{E}_{eff}(t)-\widetilde{E}_{inc}(t)=\int^T_{0}C\delta^{-\alpha}\Phi(t,\,s)\widetilde{E}_{eff}(s)ds,\quad t\in(-\infty,\,\infty),
\end{equation}
where $\widetilde{E}_{eff}$ is the effective field for $\widetilde{E}$ in $\mathbb{R}$ and $\alpha=-1+h+l$.
The differential equation with the operator $(\partial_t^2+\kappa^2)$ corresponding to the integral equation \eqref{eff-ori-system} is
\begin{equation}\label{eff-equation}
\partial_t^2\widetilde{E}_{eff}(t)+\kappa^2\widetilde{E}_{eff}(t)=-\left[C\delta^{-\alpha}\right]\chi_{[0,\,T]}\widetilde{E}_{eff}(t), \quad t\in(-\infty,\,\infty).
\end{equation}
Let $\lambda:=\sqrt{C\delta^{-\alpha}+\kappa^2}$ with
\begin{equation*}
	\lambda=\begin{cases}
		O(1), \quad &\mbox{if}\  1-h-l\geq0, \\
		O(\delta^{-\frac{\alpha}{2}}), \quad &\mbox{if}\ 1-h-l<0.
	\end{cases}\\
\end{equation*}
The general solution of the equation \eqref{eff-equation} is
\begin{equation}\label{ODE-solution}
\widetilde{E}_{eff}(t)=
\begin{cases}
	C_1\,e^{i\kappa t}+C_2\,e^{-i\kappa t},\quad & t\in(-\infty,\,0),\\
	C_3\,e^{i\lambda t}+C_4\,e^{-i\lambda t},\quad & t\in[0,\,T],\\
	C_5\,e^{i\kappa t}+C_6\,e^{-i\kappa t},\quad & t\in(T,\,\infty).
\end{cases}
\end{equation}
Setting $C_1=1$ and $C_6=0$ to satisfy the long-time condition \eqref{SRC}, the differential equation \eqref{eff-equation} with the transmission conditions  
\begin{eqnarray*}\label{transmission conditions}
&&\widetilde{E}_{eff}|_{t=0^-}=\widetilde{E}_{eff}|_{t=0^+},\\
&&\partial_t\widetilde{E}_{eff}|_{t=0^-}=\partial_t\widetilde{E}_{eff}|_{t=0^+},\\
&&\widetilde{E}_{eff}|_{t=T^-}=\widetilde{E}_{eff}|_{t=T^+},\\
&&\partial_t\widetilde{E}_{eff}|_{t=T^-}=\partial_t\widetilde{E}_{eff}|_{t=T^+},
\end{eqnarray*}
has the particular solutions with the coefficients and
\begin{eqnarray*}\label{coefficient+}
C_2&=&\frac{(\lambda^2-\kappa^2)(e^{i\lambda T}-e^{-i\lambda T})} {(\lambda+\kappa)^2e^{-i\lambda T}-(\lambda-\kappa)^2e^{i\lambda T}} =-1+\frac{4\lambda\kappa-i4\kappa^2\tan(\lambda T)}{4\lambda\kappa-i2(\lambda^2+\kappa^2)\tan(\lambda T)} \\
&=&\begin{cases}
	0+o(1), \footnotemark[4]\quad &\mbox{if $\lambda T$ is close to $n\pi$,} \\
	O(1), \quad &\mbox{if $\lambda T$ is away from $n\pi$ and $1-h-l\geq0$,}\\
	-1+O(\delta^{\frac{\alpha}{2}}), \quad &\mbox{if $\lambda T$ is away from $n\pi$ and $1-h-l<0$,}
\end{cases}\\
C_3&=&\frac{2\kappa(\kappa+\lambda)e^{-i\lambda T}} {(\lambda+\kappa)^2e^{-i\lambda T}-(\lambda-\kappa)^2e^{i\lambda T}} =\frac{2\kappa^2+2\kappa\lambda-i2\kappa(\kappa+\lambda)\tan(\lambda T)}{4\lambda\kappa-i2(\lambda^2+\kappa^2)\tan(\lambda T)}\\
&=&\frac{1}{2}+\frac{2\kappa^2+i(\lambda^2-\kappa^2-2\kappa\lambda)\tan(\lambda T)}{4\lambda\kappa-i2(\lambda^2+\kappa^2)\tan(\lambda T)} 
=\begin{cases}
	1+o(1), \quad &\mbox{if $\lambda T$ is close to $n\pi$ and $1-h-l\geq0$,} \\
	\frac{1}{2}+O(\delta^{\frac{\alpha}{2}}), \quad &\mbox{if $\lambda T$ is close to $n\pi$ and $1-h-l<0$,} \\
	O(1), \quad &\mbox{if $\lambda T$ is away from $n\pi$ and $1-h-l\geq0$,}\\
	O(\delta^{\frac{\alpha}{2}}), \quad &\mbox{if $\lambda T$ is away from $n\pi$ and $1-h-l<0$,}
\end{cases}\\
C_4&=&\frac{2\kappa(\lambda-\kappa)e^{i\lambda T}} {(\lambda+\kappa)^2e^{-i\lambda T}-(\lambda-\kappa)^2e^{i\lambda T}} =\frac{-2\kappa^2+2\kappa\lambda+i2\kappa(\lambda-\kappa)\tan(\lambda T)}{4\lambda\kappa-i2(\lambda^2+\kappa^2)\tan(\lambda T)}\\
&=&\frac{1}{2}+\frac{-2\kappa^2+i(\lambda^2-\kappa^2+2\kappa\lambda)\tan(\lambda T)}{4\lambda\kappa-i2(\lambda^2+\kappa^2)\tan(\lambda T)} 
=\begin{cases}
		0+o(1), \quad &\mbox{if $\lambda T$ is close to $n\pi$ and $1-h-l\geq0$,} \\
		\frac{1}{2}+O(\delta^{\frac{\alpha}{2}}), \quad &\mbox{if $\lambda T$ is close to $n\pi$ and $1-h-l<0$,} \\
		O(1), \quad &\mbox{if $\lambda T$ is away from $n\pi$ and $1-h-l\geq0$,}\\
		O(\delta^{\frac{\alpha}{2}}), \quad &\mbox{if $\lambda T$ is away from $n\pi$ and $1-h-l<0$,}
\end{cases}\\
C_5&=&\frac{4\lambda\kappa e^{-i\kappa T}}{(\lambda+\kappa)^2e^{-i\lambda T}-(\lambda-\kappa)^2e^{i\lambda T}} =\frac{4\lambda\kappa e^{-i\kappa T}}{4\lambda\kappa\cos(\lambda T)+i2(\lambda^2+\kappa^2)\sin(\lambda T)}\\
&=&\begin{cases}
	\pm e^{-i\kappa T}+o(1),\ \mbox{($\pm$ for even or odd $n$)}, \quad &\mbox{if $\lambda T$ is close to $n\pi$,} \\
	O(1)e^{-i\kappa T}, \quad &\mbox{if $\lambda T$ is away from $n\pi$ and $1-h-l\geq0$,}\\
	O(\delta^{\frac{\alpha}{2}})e^{-i\kappa T}, \quad &\mbox{if $\lambda T$ is away from $n\pi$ and $1-h-l<0$,}
\end{cases}
\end{eqnarray*}
where $n\in\mathbb{N}$, $n\gg1$.
\footnotetext[4]{For instance, $\kappa aT=m\pi+o(m^{-1})$.}
\medskip

Let us now prove the following theorem.
\begin{theorem}\label{effective-result-}
We have the effective field $\widetilde{E}_{eff}$ with the estimate 
\begin{equation*}
	\widetilde{E}(t)-\widetilde{E}_{eff}(t)=\begin{cases}
		O(\delta^{\frac{1-h+2l}{2}})+O(\delta^{1-h}),\quad &\mbox{if $1-h-l\geq0$,}\\
		O(\delta^{2-2h-l}), \quad &\mbox{if $\lambda T$ is close to $n\pi$, $1-h-l<0$ and $l>2\alpha$},\\
		O(\delta^{\frac{3-3h-l}{2}}), \quad &\mbox{if $\lambda T$ is away from $n\pi$, $1-h-l<0$ and $l>\frac{3}{2}\alpha$}.
	\end{cases}
\end{equation*}
\end{theorem}
{\bf Proof.}
For $m=1,\,\dots\,N$, the integral equation \eqref{eff-ori-system} at the point $t=T_m$ can be rewritten as
\begin{eqnarray}
&&\widetilde{E}_{eff}(T_m) -\frac{iC\delta^{-\alpha}}{2\kappa}\int_0^T\sum_{j=1}^{N}\widetilde{\Phi}(T_m,\,T_j)\widetilde{E}_{eff}(T_j)x_jds\nonumber\\
&&=\widetilde{E}_{inc}(T_m)+\sum_{j=0}^{N}\int^{T_{j+1}}_{T_j}\frac{iC\delta^{-\alpha}}{2\kappa}\left[\widetilde{\Phi}(T_m,\,s)\widetilde{E}_{eff}(s)-\widetilde{\Phi}(T_m,\,T_j)\widetilde{E}_{eff}(T_j)\right]ds\nonumber\\
&&\qquad\qquad\quad\; +\frac{iC\delta^{-\alpha}}{2\kappa}\int^{T_{1}}_{T_0}\widetilde{\Phi}(T_m,\,T_0)\widetilde{E}_{eff}(T_0)ds\label{eff-f-1}.
\end{eqnarray}
Denote $f_j(T_m,\,s):=\widetilde{\Phi}(T_m,\,s)\widetilde{E}_{eff}(s)$ for $s\in [T_j,\,T_{j+1}],\ j=0,\,1,\,2,\,\dots,\,N$.
With the Taylor series, we have
\begin{equation*}
f_j(T_m,\,s)-f_j(T_m,\,T_j)=(s-T_j)\bar{R}_j(T_m,\,s), 
\end{equation*}
where 
\begin{eqnarray*}
\bar{R}_j(T_m,s) &:=&\int_0^1\partial_sf(T_m,\,s-\vartheta(s-T_j))\,d\vartheta\\
&=&\int_0^1\left[\partial_s\widetilde{\Phi}(T_m,\,s-\vartheta(s-T_j))\right]\widetilde{E}_{eff}(s-\vartheta(s-T_j))\,d\vartheta\\
&&+\int_0^1\widetilde{\Phi}(T_m,\,s-\vartheta(s-T_j))\left[\partial_s\widetilde{E}_{eff}(s-\vartheta(s-T_j))\right]d\vartheta, \end{eqnarray*}
and  
\begin{eqnarray*}
|\bar{R}_j(T_m,s)| &\leq&\Bigg|\int_0^1i\kappa(\vartheta-1)\widetilde{\Phi}(T_m,\,s-\vartheta(s-T_j))\widetilde{E}_{eff}(s-\vartheta(s-T_j))\,d\vartheta\Bigg|\\
&&+\Bigg|\int_0^1\widetilde{\Phi}(T_m,\,s-\vartheta(s-T_j))\left[\partial_s\widetilde{E}_{eff}(s-\vartheta(s-T_j))\right]d\vartheta\Bigg|\nonumber\\
&\leq&\kappa\left(\int_0^1|\vartheta-1|^2d\vartheta\right)^{\frac{1}{2}}||\widetilde{E}_{eff}||_{L^2[T_j,\,T_{j+1}]} +||\partial_t\widetilde{E}_{eff}||_{L^2[T_j,\,T_{j+1}]}\\
&=&\begin{cases}
O(\delta^{\frac{l+\alpha}{2}})+O(\delta^{\frac{l}{2}}), \quad &\mbox{if $\lambda T$ is away from $n\pi$ and $1-h-l<0$,} \\
O(\delta^{\frac{l}{2}})+O(\delta^{\frac{l-\alpha}{2}}), \quad &\mbox{if $\lambda T$ is close to $n\pi$ and $1-h-l<0$,} \\
O(\delta^{\frac{l}{2}}), \quad &\mbox{otherwise.}
\end{cases}\\
&=&\begin{cases}
	O(\delta^{\frac{1-h}{2}}), \quad &\mbox{if $\lambda T$ is close to $n\pi$ and $1-h-l<0$,} \\
	O(\delta^{\frac{l}{2}}), \quad &\mbox{otherwise.}
\end{cases}
\end{eqnarray*}
Then, the estimates of the integrals in the right of \eqref{eff-f-1} are 
\begin{eqnarray*}
|I_A|&:=&\Bigg|\sum_{j=0}^{N}\int^{T_{j+1}}_{T_j}\frac{iC\delta^{-\alpha}}{2\kappa}\left[\widetilde{\Phi}(T_m,\,s)\widetilde{E}_{eff}(s)-\widetilde{\Phi}(T_m,\,T_j)\widetilde{E}_{eff}(T_j)\right]ds\Bigg|\nonumber\\
&\leq&\sum_{j=0}^{N}\int^{T_{j+1}}_{T_j} \frac{iC\delta^{-\alpha}}{2\kappa}|s-T_j|\,|\bar{R}_j(T_m,s)|ds\\
&=&\begin{cases}
	O(\delta^{\frac{2-2h+l}{2}}), \quad &\mbox{if $\lambda T$ is close to $n\pi$ and $1-h-l<0$,} \\
	O(\delta^{\frac{1-h+2l}{2}}), \quad &\mbox{otherwise,}
\end{cases}
\end{eqnarray*} 
and
\begin{eqnarray*}
	|I_B|&:=&\Bigg|\frac{iC\delta^{-\alpha}}{2\kappa}\int^{T_{1}}_{T_0}\widetilde{\Phi}(T_m,\,T_0)\widetilde{E}_{eff}(T_0)ds\Bigg|\nonumber\\
	&=&O(\delta^{l-\alpha}(C_3+C_4))=\begin{cases}
		O(\delta^{\frac{1-h+l}{2}}), \quad &\mbox{if $\lambda T$ is away from $n\pi$ and $1-h-l<0$,} \\
		O(\delta^{1-h}), \quad &\mbox{otherwise,}
	\end{cases}
\end{eqnarray*} 
Thus, \eqref{eff-f-1} can be rewritten as
\begin{equation}\label{EFF-est}
\widetilde{E}_{eff}(T_m) -\frac{iC\delta^{-\alpha}}{2\kappa}\int_0^T\sum_{j=1}^{N}\widetilde{\Phi}(T_m,\,T_j)\widetilde{E}_{eff}(T_j)x_jds
=\widetilde{E}_{inc}(T_m)+|I_A|+|I_B|,
\end{equation}
with 
\begin{equation*}
	|I_A|+|I_B|=\begin{cases}
		O(\delta^{\frac{1-h+2l}{2}})+O(\delta^{1-h}),\quad &\mbox{if $1-h-l\geq0$,}\\
		O(\delta^{1-h}), \quad &\mbox{if $\lambda T$ is close to $n\pi$ and $1-h-l<0$,}\\
		O(\delta^{\frac{1-h+l}{2}}), \quad &\mbox{if $\lambda T$ is away from $n\pi$ and $1-h-l<0$.}
	\end{cases}
\end{equation*}

Denote $Y_m:=\delta^{-1}\int^{T_m+\delta/2}_{T_m-\delta/2}\widetilde{E}(t)dt$, then according to \eqref{try-multiple-m} and \eqref{GG1}, we have
\begin{equation}\label{EFF-alg-system-with-error}
Y_m -\frac{iC\delta^{-\alpha}}{2\kappa} \int_0^T\sum_{j=1}^{N} \widetilde{\Phi}(T_m,\,T_j)x_jY_jds =\delta^{-1}\int^{T_m+\delta/2}_{T_m-\delta/2}\widetilde{E}_{inc}(t)\,dt +O\left(\delta^{2-h-l}\right).
\end{equation}
Let the algebraic system of \eqref{EFF-alg-system-with-error} be
\begin{equation}\label{EFF-alg-system-without-error}
\hat{Y}_m -\frac{iC\delta^{-\alpha}}{2\kappa} \int_0^T\sum_{j=1}^{N} \widetilde{\Phi}(T_m,\,T_j)\hat{Y}_jx_jds =\delta^{-1}\int^{T_m+\delta/2}_{T_m-\delta/2}\widetilde{E}_{inc}(t)\,dt.
\end{equation}
Taking the difference between \eqref{EFF-alg-system-without-error} and \eqref{EFF-est}, we can get the algebraic system
\begin{equation*}
(\hat{Y}_m-\widetilde{E}_{eff}(T_m)) -\frac{iC\delta^{-\alpha}}{2\kappa}\int_0^T\sum_{j=1}^{N}\widetilde{\Phi}(T_m,\,T_j)(\hat{Y}_j-\widetilde{E}_{eff}(T_j))x_jds
=O(\kappa\delta)+|I_A|+|I_B|,
\end{equation*}
since $\delta^{-1}\int^{T_m+\delta/2}_{T_m-\delta/2}\widetilde{E}_{inc}(t)dt-\widetilde{E}_{inc}(T_m)=O(\kappa\delta)$. Denote $Z_m:=\hat{Y}_m-\widetilde{E}_{eff}(T_m)$ and $Z:=(Z_1,\,Z_2,\,\dots,\,Z_N)^T$.
According to the estimate of the algebraic system \eqref{EFF-alg-system-without-error}, we have 
\begin{eqnarray*}
||Z||_2 &=&||A^{-1}||_2(O(\kappa\delta)+|I_A|+|I_B|)||\widetilde{e}||_2\\
&=&\begin{cases}
	O(\delta^{\frac{1-h+l}{2}})+O(\delta^{\frac{2-2h-l}{2}}),\quad &\mbox{if $1-h-l\geq0$,}\\
	O(\delta^{\frac{2-2h-l}{2}}), \quad &\mbox{if $\lambda T$ is close to $n\pi$ and $1-h-l<0$,}\\
	O(\delta^{\frac{1-h}{2}}), \quad &\mbox{if $\lambda T$ is away from $n\pi$ and $1-h-l<0$.}
\end{cases}
\end{eqnarray*}

For $t\in\mathbb{R}$, according to \eqref{try-multiple-1-3}, we have
\begin{equation}\label{Y_m}
\widetilde{E}(t)-\widetilde{E}_{inc}(t) =\frac{iC\delta^{1-h}}{2\kappa}\sum_{j=1}^N \widetilde{\Phi}(t,\,T_j)\hat{Y}_j +O(\delta^{2-h-l}).
\end{equation}
Taking the difference between \eqref{Y_m} and \eqref{eff-ori-system}, we get 
\begin{eqnarray*}
\widetilde{E}_{eff}(t)-\widetilde{E}(t)&=&\int^T_{0}\frac{iC\delta^{-\alpha}}{2\kappa}\widetilde{\Phi}(t,\,s)\widetilde{E}_{eff}(s)\,ds-\frac{iC\delta^{1-h}}{2\kappa}\sum_{j=1}^N \widetilde{\Phi}(t,\,T_j)\hat{Y}_j +O(\delta^{2-h-l})\nonumber\\
&=&\sum_{j=0}^{N}\int^{T_{j+1}}_{T_j}\frac{iC\delta^{-\alpha}}{2\kappa}\left[\widetilde{\Phi}(t,\,s)\widetilde{E}_{eff}(s)-\widetilde{\Phi}(t,\,T_j)\widetilde{E}_{eff}(T_j)\right]ds\nonumber\\
&&+\frac{iC\delta^{-\alpha}}{2\kappa}\int^{T_{1}}_{T_0}\widetilde{\Phi}(T_m,\,T_0)\widetilde{E}_{eff}(T_0)ds\\
&&+\frac{iC\delta^{1-h}}{2\kappa}\sum_{j=1}^{N}\widetilde{\Phi}(t,\,T_j)\left[\widetilde{E}_{eff}(T_j)-\hat{Y}_j\right] +O(\delta^{2-h-l})\nonumber\\
&\leq&|I_A|+|I_B|+O(\delta^{1-h})\left(\sum_{j=1}^{N}|\widetilde{\Phi}(t,\,T_j)|^2\right)^{\frac{1}{2}}\left(\sum_{j=1}^{N}|\widetilde{E}_{eff}(T_j)-\hat{Y}_j|^2\right)^{\frac{1}{2}}+O(\delta^{2-h-l})\nonumber\\
&=&|I_A|+|I_B|+O(\delta^{1-h})\,||\hat{F}||_2||Z||_2+O(\delta^{2-h-l})\nonumber\\
&=&\begin{cases}
	O(\delta^{\frac{1-h+2l}{2}})+O(\delta^{1-h}),\quad &\mbox{if $1-h-l\geq0$,}\\
	O(\delta^{2-2h-l}), \quad &\mbox{if $\lambda T$ is close to $n\pi$, $1-h-l<0$ and $l>2\alpha$},\\
	O(\delta^{\frac{3-3h-l}{2}}), \quad &\mbox{if $\lambda T$ is away from $n\pi$, $1-h-l<0$ and $l>\frac{3}{2}\alpha$}.
\end{cases}
\end{eqnarray*}
With $\alpha=-1+h+l$, 
\begin{equation*}
	\widetilde{E}_{eff}(t)-\widetilde{E}(t)=\begin{cases}
		O(\delta^{\frac{1-h+2l}{2}})+O(\delta^{1-h}),\quad &\mbox{if $1-h-l\geq0$,}\\
		O(\delta^{2-2h-l}), \quad &\mbox{if $\lambda T$ is close to $n\pi$, $1-h<l<2(1-h)$},\\
		O(\delta^{\frac{3-3h-l}{2}}), \quad &\mbox{if $\lambda T$ is away from $n\pi$, $1-h<l<3(1-h)$}.
	\end{cases}
\end{equation*}
The proof is complete.\hfill $\Box$

\section{Numerical examples}\label{Numerical-tests}
In this section, we give several numerical examples to verify the results provided in Theorem \ref{effective-result} and illustrate the accuracy of the effective field. We compare the effective field with the asymptotically approximated field given through the algebraic system. 

\begin{example}\label{exam2}
	Consider the model \eqref{maxwell-D-Dinc-kperp} based on Definition 1.1. We test the error of the effective field $\tilde{E}_{eff}$ as \eqref{effective-field} and the asymptotic field $\tilde{E}$ as \eqref{result-equation} respectively.
\end{example}

\begin{figure}[htp]
	\begin{center}
		\includegraphics[width=0.4\textwidth,height=4.5cm]{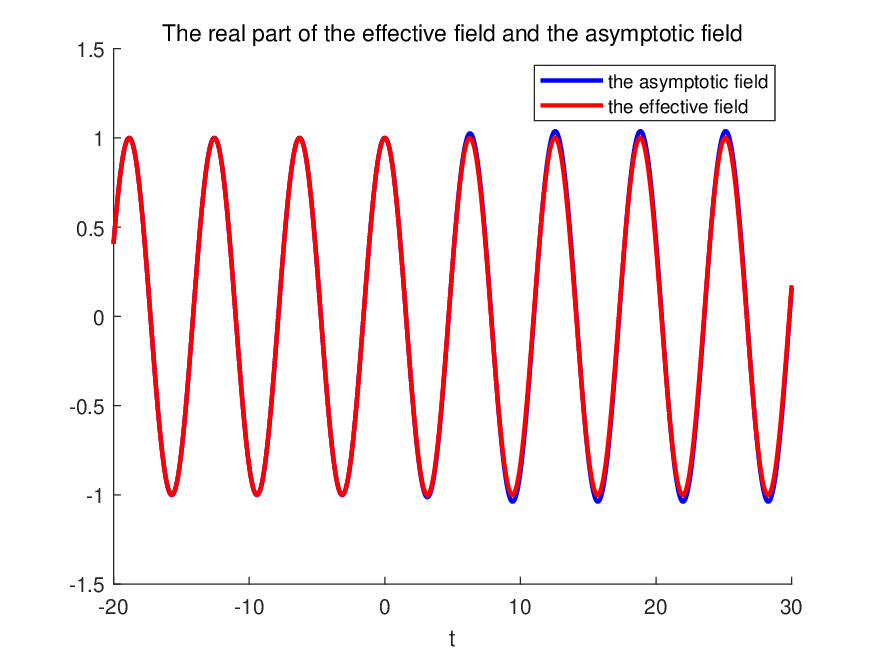}
		\includegraphics[width=0.4\textwidth,height=4.5cm]{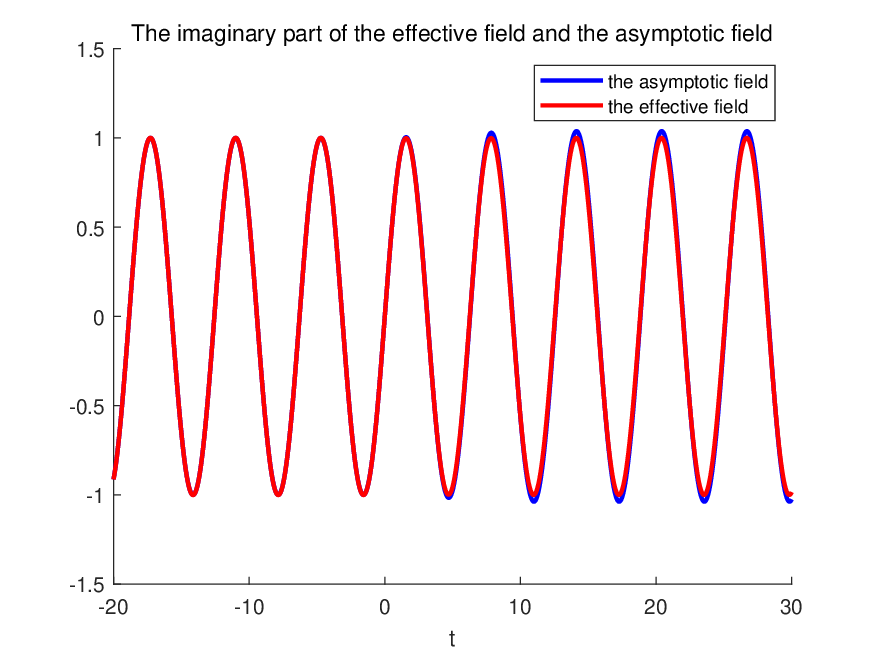}

		\includegraphics[width=0.4\textwidth,height=4.5cm]{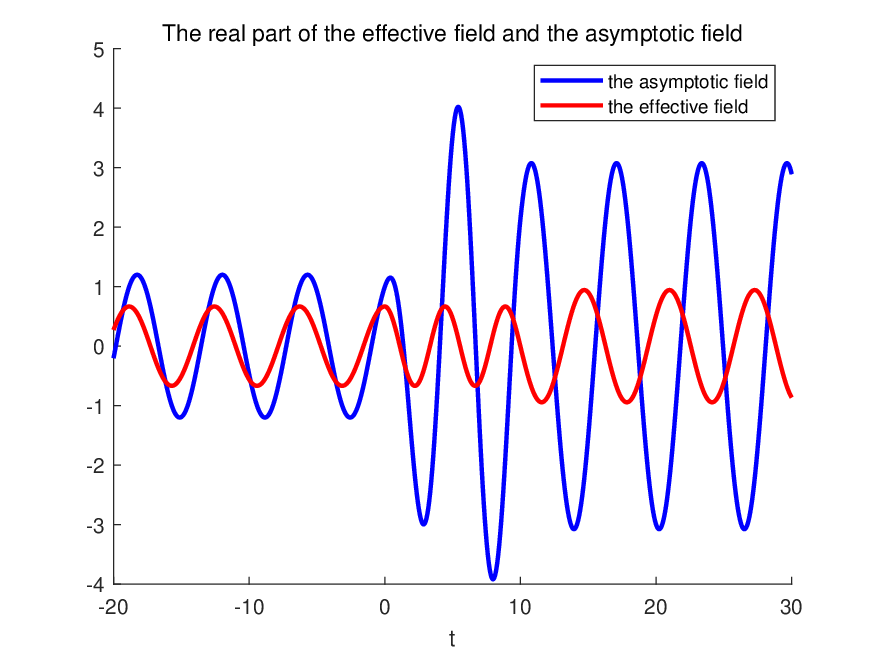}
		\includegraphics[width=0.4\textwidth,height=4.5cm]{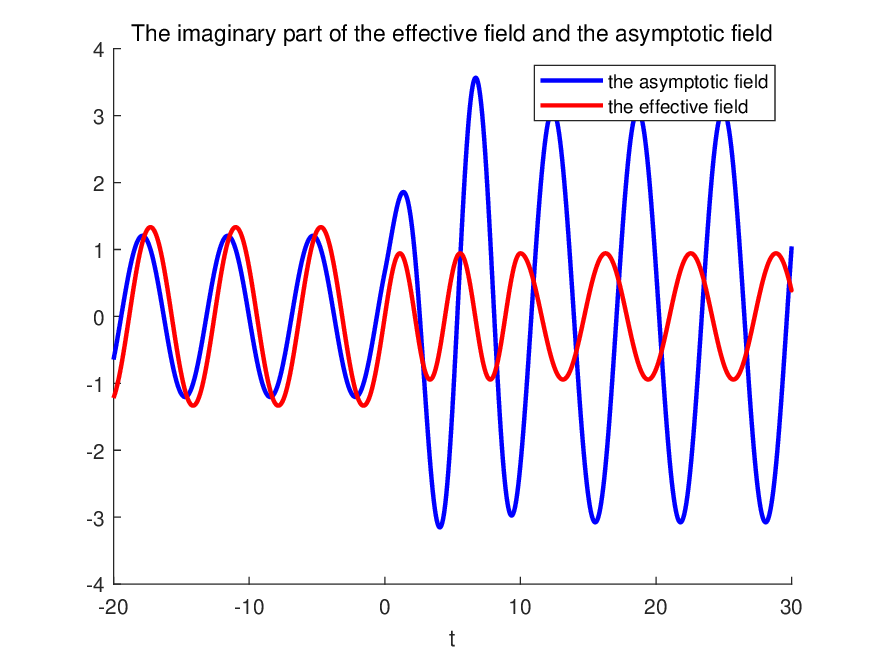}

		\includegraphics[width=0.4\textwidth,height=4.5cm]{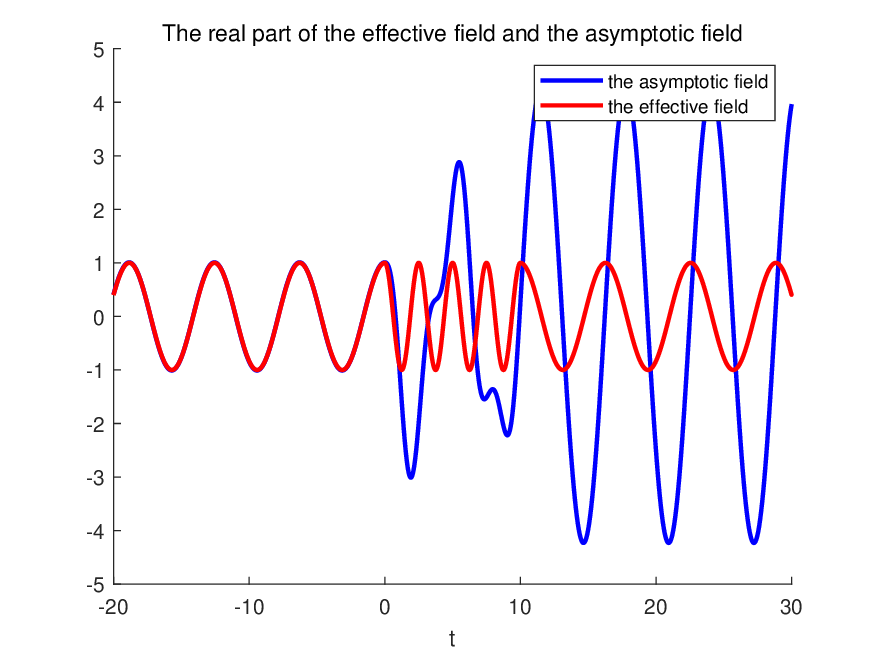}
		\includegraphics[width=0.4\textwidth,height=4.5cm]{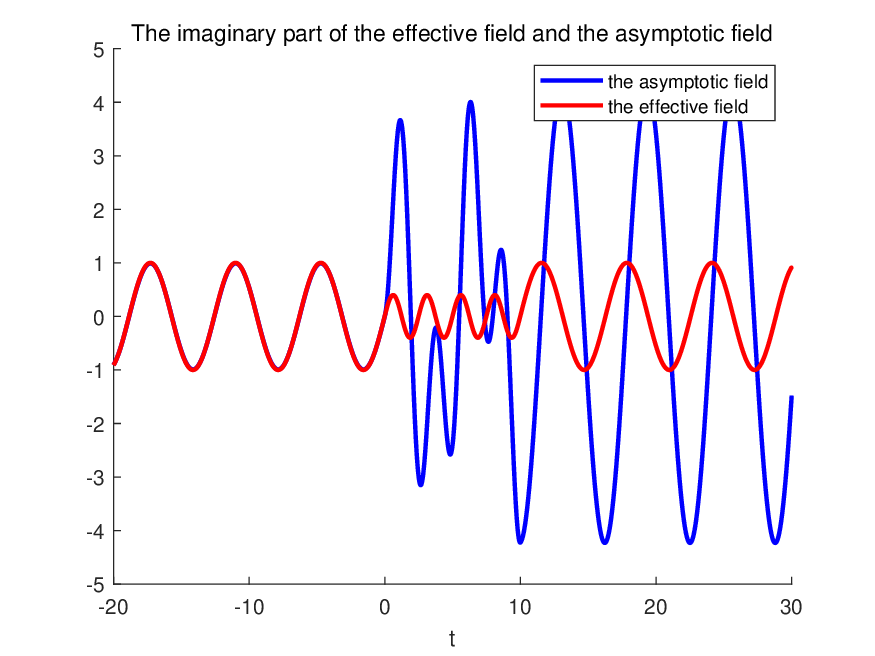}
		\caption{Numerical results for the effective field $\tilde{D}_{eff}$ and the asymptotic field: when $1-h-l>0$ (upper),\quad $1-h-l=0$ (middle),\quad $1-h-l<0$ (lower).}\label{exam2-p-1}
	\end{center}
\end{figure}

In Table 2, we list the parameter settings according to Definition \ref{defn1} for different values of $h$ and $l$ in Example \ref{exam2}. The numerical results of the asymptotic field and the effective field can be found in Figures \ref{exam2-p-1}. 

From the first line in Figure \ref{exam2-p-1}, as the case of $1-h-l>0$, the difference between the asymptotic field and the effective field is small. In addition, it verifies that both the asymptotic field and the effective field are close to the incident wave. 
However, it can be observed from other lines in Figure \ref{exam2-p-1} that when $1-h-l\leq0$, the behaviors of the asymptotic field and the effective field are different. For $t<0$, these two fields are basically consistent. But for $t\geq 0$, the asymptotic field is no longer reliable because of the ill-posedness of the inverse of the  matrix $\bar{A}$. This shows and confirms that the effective field is a better choice, for the case of $1-h-l\geq0$, to simulate the wave fields generated by such (many) inhomogeneities.

\begin{center}
	\textbf{Table 2.} The parameter settings in Example \ref{exam2}\\[+1mm]
	\begin{tabular}{cccccccc}\hline\hline\\[-3.5mm]
		$T$ & $h$ & $l$ &$\delta$ & $\kappa$ & $C$ & $\omega_p^2$ & $\lambda$ \\ \hline
		
		10 & 0.1 & 0.1 & 1e-03 & 1 & 1 & 1.9953 & 1.0020 \\
		
		10 & 0.1 & 0.9 & 1e-03 & 1 & 1 & 1.9953 &  1.4142 \\
		
		10 & 0.342 & 0.9 & 1e-03 & 1 & 1 & 10.6170 & 2.5142 \\
		\hline\hline
	\end{tabular}
\end{center}

\bigskip
{\bf Acknowledgement:} This work is done while the third author Q. Yao was visiting the Radon Institute, Linz (Austria). She is supported by National Natural Science Foundation of China (No. 12071072), China Scholarships Council (No.202006090298) and the Startup Foundation for Introducing Talent of NUIST. The first author M. Sini is partially supported by the Austrian Science Fund FWF: (FWF): P 32660. The second author H. Wang is partially supported by National Natural Science Foundation of China (No. 12071072).


\begin{thebibliography}{99}

\bibitem{ACP-2}
H. Ammari, D.P. Challa, A.P. Choudhury, M. Sini, The equivalent media generated by bubbles of high contrasts: volumetric metamaterials and metasurfaces, Multiscale Model. Simul., 18 (2020), 240--293.

\bibitem{AFLYH}
H. Ammari, B. Fitzpatrick, H. Lee, S. Yu, H. Zhang, Double-negative acoustic metamaterials, Quart. Appl. Math., 77 (2019), 767--791.

\bibitem{AH}
H. Ammari, H. Zhang, Effective medium theory for acoustic waves in bubbly fluids near Minnaert resonant frequency, SIAM J. Math. Anal., 49 (2017), 3252--3276.

\bibitem{Bensoussan}
A. Bensoussan, J.-L. Lions, G. Papanicolaou, Asymptotic Analysis for Periodic Structures, AMS Chelsea Publishing, Providence, RI, 2011.

\bibitem{Bouchitte-Schweizer}
G. Bouchitt\'e, B. Schweizer, Homogenization of Maxwell's equations in a split ring geometry, Multiscale Model. Simul., 8 (2010), 717--750.

\bibitem{BBF}
G. Bouchitt\'e, C. Bourel, D. Felbacq, Homogenization near resonances and artificial magnetism in three dimensional dielectric metamaterials, Arch. Ration. Mech. Anal., 225 (2017), 1233--1277.

\bibitem{AM-1}
A. Bouzekri, M. Sini, The Foldy-Lax approximation for the full electromagnetic scattering by small conductive bodies of arbitrary shapes, Multiscale Model. Simul., 17 (2019), 344--398.

\bibitem{AM}
A. Bouzekri, M. Sini, Mesoscale approximation of the electromagnetic fields, Ann. Henri Poincar\'{e}, 22 (2021), 1979--2028.

\bibitem{Caloz-Leger-2020}
C. Caloz, Z. Deck-L\'eger, Spacetime metamaterials--part I: general concepts, IEEE Trans. Antennas Propag., 68 (2020), 1569--1582.

\bibitem{Challa-1}
D.P. Challa, M. Andrea, M. Sini, Characterization of the acoustic fields scattered by a cluster of small holes, Asymptot. Anal., 118 (2020), 235--268.

\bibitem{Chen-Lipton-2013}
Y. Chen, R. Lipton, Resonance and double negative behavior in metamaterials, Arch. Ration. Mech. Anal., 209 (2013), 835--868.

\bibitem{Cher-Cooper}
K. Cherednichenko, S. Cooper, Homogenization of the system of high-contrast Maxwell equations, Mathematika, 61 (2015), 475--500.

\bibitem{Cher-Ersh-Kise}
K.D. Cherednichenko, Y.Y. Ershova, A.V. Kiselev, Effective behaviour of critical-contrast PDEs: micro-resonances, frequency conversion, and time dispersive properties. I., Comm. Math. Phys., 375 (2020), 1833--1884. 

\bibitem{Cioranescu-Murat}
D. Cioranescu, F. Murat, A strange term coming from nowhere, Topics in the Mathematical Modelling of Composite Materials, Birkh\"{a}user, Boston, MA, 1997.  

\bibitem{Engheta:2013}
N. Engheta, Pursuing near-zero response, Science, 340 (2013), 286--287.

\bibitem{Galiffi-al-2022}
E. Galiffi, R. Tirole, S. Yin, H. Li, S. Vezzoli, P.A. Huidobro, M.G. Silveirinha, R. Sapienza, A. Alu, J.B. Pendry, Photonics of time-varying media, Adv. Photonics, 4 (2022), 014002. https://doi.org/10.1117/1.AP.4.1.014002

\bibitem{Huidobro-et-al}
P.A. Huidobro, E. Galiffi, S. Guenneau, R.V. Craster, J.B. Pendry, Fresnel drag in space-time-modulated metamaterials, Proc. Natl. Acad. Sci. USA., 116 (2019), 24943--24948.

\bibitem{Huidobro--Pendry} 
P.A. Huidobro, M.G. Silverinha, E. Galiffi, J.B. Pendry, Homogenisation theory of space-time metamaterials, Phys. Rev. Appl., 16 (2021), 014044.


\bibitem{Jikov}
V.V. Jikov, S.M. Kozlov, O.A. Oleinik, Homogenization of Differential Operators and Integral Functionals, Springer-Verlag, Berlin, 1994.

\bibitem{Kalluri-book}
D.K. Kalluri, Electromagnetics of Time Varying Complex Media Frequency and Polarization transformer, Second Edition, CRC Press, Boca Raton, 2010.

\bibitem{L-S:2016}
A. Lamacz, B. Schweizer, A negative index meta-material for Maxwell's equations, SIAM J. Math. Anal., 48 (2016), 4155--4174.

\bibitem{Lannebere-Morgado-Silveirinha} 
S. Lanneb\'ere, T.A. Morgado, M.G. Silveirinha, First principles homogenization of periodic metamaterials and application to wire media, arXiv: 2002.06271.

\bibitem{Lurie} 
K.A. Lurie, An Introduction to the Mathematical Theory of Dynamic Materials, Springer, Cham, 2017.

\bibitem{Mazya-1} 
V.G. Maz'ya, A.B. Movchan, M.J. Nieves, Green's Kernels and Meso-Scale Approximations in Perforated Domains, Springer, Heidelberg, 2013.


\bibitem{Mazya-16}
V.G. Maz'ya, A.B. Movchan, M.J. Nieves, Mesoscale models and approximate solutions for solids containing clouds of voids, Multiscale Model. Simul., 14 (2016), 138--172.


\bibitem{Nieves-17}
M.J. Nieves, Asymptotic analysis of solutions to transmission problems in solids with many inclusions, SIAM J. Appl. Math., 77 (2017), 1417--1443.

\bibitem{Pena}
V. Pacheco-Pe\~{n}a, N. Engheta, Temporal aiming, Light Sci. Appl., 9 (2020), 129.


\bibitem{Ramm-7}
A.G. Ramm, Many-body wave scattering by small bodies and applications, J. Math. Phys., 48 (2007), 103511.


\bibitem{Ramm-15}
A.G. Ramm, Scattering of electromagnetic waves by many small perfectly conducting or impedance bodies, J. Math. Phys., 56 (2015), 091901.

\bibitem{Schweizer:2017}
B. Schweizer, Resonance meets homogenization: construction of meta-materials with astonishing properties,  Jahresber. Dtsch. Math.-Ver., 119 (2017), 31--51.

\bibitem{Sini-Wang-1}
M. Sini, H. Wang, Q. Yao, Analysis of the acoustic waves reflected by a cluster of small holes in the time-domain and the equivalent mass density, Multiscale Model. Simul., 19 (2021), 1083--1114.

\bibitem{Suslina-2019}
T.A. Suslina, Homogenization of the stationary Maxwell system with periodic coefficients in a bounded domain, Arch. Ration. Mech. Anal., 234 (2019), 453--507.

\end{thebibliography}
\end{document}